\documentclass[11pt,a4paper,final]{article}
\usepackage{ifpdf}               % This loads the \ifpdf \else \fi structure
\usepackage{a4wide}             % A wider use of a4paper
\usepackage{amssymb}            % Some Useful symbols
\usepackage{slashed}            % For Feynman slashes
\usepackage[latin1]{inputenc}    % This allows you to write é in stead of \'e
\usepackage[notref,notcite]{showkeys} 
\ifpdf
\usepackage[pdftex]{graphicx}
\usepackage[pdftex,unicode,implicit]{hyperref}
\hypersetup{%
  pdftitle    = {Fake supersymmetric solutions in supergravity},
  pdfkeywords = {Fake supersymmetry, black holes, Kasner Universes, Nariai cosmos},
  pdfauthor   = {P. Meessen},
  pdfcreator  = {pdf\LaTeXe\ with package \flqq hyperref\frqq},
  pdfproducer = {pdf\LaTeXe\ with package \flqq hyperref\frqq},
  pdfpagemode = None,  %change this to FullScreen if you want...
  pdffitwindow= true,  % Open the pdf file without the sidebars open.
  unicode     = true,
  plainpages  = false,
  colorlinks  = true,  % links have no different colors..
  citecolor   = black,
  urlcolor    = black,
  linkcolor   = black
}

\else
  \usepackage[dvips]{graphicx}
  \usepackage[unicode,implicit]{hyperref}

\fi
\begin{document}
%\begin{center}
%  {\bf Fake De Sitter supergravity coupled to vector multiplets}
%\end{center}
\begin{flushright}
\small
IFT-UAM/CSIC-09-08\\
%{\bf arXiv:yymm.nnnn}\\
%\today
27 February 2009
\normalsize
\end{flushright}
\begin{center}
{\large\bf Cosmological solutions from fake $N=2$ EYM supergravity}\\[.5cm]
{\bf Patrick~Meessen and Alberto~Palomo-Lozano}\\[.2cm]%\footnote{E-mail: {\tt patrick.meessen@uam.es}}\\[.2cm]
{\em Instituto de F\'{\i}sica Te\'orica UAM/CSIC, Facultad de Ciencias C-XVI\\
     C.U. Cantoblanco, E-28049 Madrid, Spain}\\[.5cm]
{\bf abstract}\\
\begin{quote}
{\small 
  We characterise the (fake) supersymmetric solutions of Wick-rotated $N=2$ $d=4$ gauged supergravity coupled
  to non-Abelian vector multiplets. In the time-like case we obtain generalisations of Kastor {\&} Traschen's
  cosmological black holes: 
  they have a specific time-dependence and the base-space must be 3-dimensional hyperCR/Gauduchon-Tod space.
  In the null-case, we find that the metric has a holonomy contained in $\mathrm{Sim}(2)$, give a general 
  characterisation of the solutions, and give some examples.
  Finally, we point out that in some cases the solutions we found are non-BPS solutions to $N=2$ $d=4$ supergravity
  coupled to vector multiplets. 
}
\end{quote}
\end{center}
\vspace{.5cm}
%%%%%%%%%%%%%%%%%%%%%%%%%%%%%%% End of the headers   %%%%%%%%%%%%%%%%%%%%%%%%%%%%%%%%%%%%%%%%%%%%%%%%%%%%%%%%%%%%%

%%%%%%%%%%%%%%%%%%%%%%%%%%%%%%%%%%%%%%%%%%%%%%%%%%%%%%%%%%%%%%%%%%%%%%%%%%%%%%%%%%%%%%%%%%%%%%%%%%%%%%%%%%%%%%%%%%
In 1992 Kastor {\&} Traschen \cite{Kastor:1992nn} created a cosmological multi-black hole solution to
Einstein-Maxwell-De Sitter gravity by
observing that the extreme Reissner-Nordstrom-De Sitter black hole solution written in spherical coordinates could be 
transformed to the time-dependent conforma-static form
\begin{equation}
  \label{eq:KT1}
  ds^{2} \; =\; \Omega^{-2}d\tau^{2} \ -\ \Omega^{2}\ d\vec{x}^{2}_{(3)} \hspace{.6cm}\mbox{with}\hspace{.6cm}
  \Omega \ =\ H\tau \ +\ \frac{m}{r} \; ,
\end{equation}
where $3H^{2}$ is the cosmological constant and we introduced the further coordinate transformation $H\tau =e^{Ht}$;
as the $r$-dependent part of $\Omega$ is a spherically symmetric harmonic function, the multi-bh solutions can be 
created by changing it to a general harmonic function.
\par
Seeing the similarity of the above solution and the supersymmetric solutions to minimal $N=2$ $d=4$ supergravity
\cite{Tod:1983pm},
whose bosonic part is just EM-theory, Kastor {\&} Traschen showed \cite{Kastor:1993mj}
that their multi-bh solution solved the spinorial equations\footnote{
  In this article we will be following the conventions of ref.~\cite{Hubscher:2008yz}, which in its turn are 
  adapted from those of ref.~\cite{Andrianopoli:1996cm}.
  Specifically this means that the metric is mostly-minus, the $\gamma$-matrices are purely imaginary and the 
  spinors are chiral, with $\gamma_{5}\epsilon^{I}=\epsilon^{I}$ and $\gamma_{5}\epsilon_{I}=-\epsilon_{I}$. 
  As $\gamma_{5}=-i\gamma^{0123}$ is purely imaginary, the above chirality assignment is compatible with the convention
  of raising and lowering $I$-indices by complex conjugation.
}
\begin{equation}
  \label{eq:FM6a}
  \nabla_{a}\epsilon_{I} \, =\, -\textstyle{iH\over 2}\gamma_{a}\ \varepsilon_{IJ}\epsilon^{J}
                             \ +\ H\ A_{a}\epsilon_{I}
                             \ +\ iF^{+}_{ab}\gamma^{b}\varepsilon_{IJ}\epsilon^{J}\; .
\end{equation}
This fermionic rule can be derived from the supersymmetry variations of minimal gauged $N=2$ $d=4$ supergravity, which has
an anti-De Sitter type cosmological constant $\Lambda = 3g^{2}$, by Wick-rotating $g\rightarrow iH$. 
As eq.~(\ref{eq:FM6a}) looks like a Killing Spinor Equation but is not due to supersymmetry, we will refer to
equations like it as fake-Killing Spinor Equations (fKSEs) \cite{Freedman:2003ax}.
\par
The KT solutions were subsequently generalised to higher dimensions by London in ref.~\cite{London:1995ib}, who also showed that
his solutions solved a suitable fKSE, and generalised to spinning solutions in a stringy theory\footnote{
  The model used by Shiromizu can be seen as a truncation of a model
  with prepotential $\mathcal{F}=-i/2\mathcal{X}^{0}\mathcal{X}^{1}$ and $\mathtt{C}_{1}=0$, the meaning
  of which will be explained in section (\ref{sec:DSVector}). His solutions can be obtained from the results
  in section (\ref{sec:VectBil}).
} 
by Shiromizu \cite{Shiromizu:1999xj}.
In ref.~\cite{Behrndt:2003cx}, Behrndt {\&} Cveti\v{c} generalised the KT-solution to asymptotically DS solutions to
$5$- and $4$-dimensional supergravities coupled to vector multiplets by observing the following substitution
rule: as one can see form the expression for $\Omega$ in eq.~(\ref{eq:KT1}), the difference between
the cosmological solution and the usual supersymmetric solutions is nothing but the linear $\tau$-dependence.
As harmonic functions appear quite natural in supersymmetric solutions, the substitution rule is to add to these harmonic 
functions a piece linear in the time-coordinate. Furthermore, Behrndt {\&} Cveti\v{c} showed that their solutions
solved fKSEs that could be obtained from the KSEs of gauged supergravity coupled to vector-multiplets, by 
Wick-rotating the coupling constant, pointing out that this is equivalent to considering an $\mathbb{R}$-gauged
symmetry. Indeed, the construction of {\em e.g.\/} gauged $N=2$ $d=4$ supergravity coupled to vector-multiplets
calls for the inclusion of an $U(1)$ Fayet-Iliopoulos term, which as far as the Killing spinor is concerned means
that it is gauged (see {\em e.g.\/} \cite{Andrianopoli:1996cm}), proportional to the coupling constant. Wick-rotating
the coupling constant, then, is equivalent to Wick-rotating the gauge group, which becomes $\mathbb{R}$.
\par
It was recently realised by Grover {\em et al.\/} \cite{Grover:2008jr}, that the techniques used to classify 
supersymmetric solutions to supergravity theories, could be used to construct solutions to theories admitting
fKSEs; 
they applied the techniques of ref.~\cite{Gillard:2004xq} to the classification of solutions to the time-like case of 
minimal `De Sitter $N=1$ $d=5$ supergravity', which can be obtained by Wick-rotation from 
minimal gauged $N=1$ $d=5$ supergravity. Of special interest in these classification is the geometry of the 
4-dimensional base-space, which in the DS case turns out to be hyperK\"ahler-Torsion, whereas it is hyper-K\"ahler
in the ungauged sugra \cite{Gauntlett:2002nw} and K\"ahler in the gauged sugra \cite{Gauntlett:2003fk}.
\par
In this article we will extend the results of K{\&}T \cite{Kastor:1992nn} and B{\&}C \cite{Behrndt:2003cx}
by considering solutions to Wick'ed $N=2$ $d=4$ 
supergravity coupled to non-Abelian vector multiplets, by which we mean that we allow for gaugings of 
the isometries of the scalar manifold (see {\em e.g.\/} \cite{Andrianopoli:1996cm}).
As indicated above, this theory can be obtained from 
gauged $N=2$ $d=4$ supergravity coupled to non-Abelian vector-multiplets by Wick-rotation, not of the coupling constant 
as we are allowing for non-Abelian couplings, but of the Fayet-Iliopoulos term responsible for gauging the $R$-symmetry; 
we shall refer to this theory as fake $N=2$ Einstein-Yang-Mills.
{}For understandable reasons, the $N=2$ $d=4$ supergravity theories have attracted quite some interest in the last
decades, and the theories for which the supersymmetric solutions have been fully classified/characterised are
the minimal theory \cite{Tod:1983pm}, 
the minimal theory coupled to vector- and hyper-multiplets \cite{Behrndt:1997ny,Meessen:2006tu},
the minimal theory coupled to non-Abelian vector-multiplets \cite{Hubscher:2008yz},
minimal gauged theory \cite{Caldarelli:2003pb}, 
and recently the minimal gauged theory coupled to vector-multiplets \cite{Cacciatori:2008ek}.
\par
Wick rotation of the coupling constant in gauged supergravity was also considered in ref.~\cite{Skenderis:2007sm} 
in order to
find a supergravity basis for the Domain Wall/Cosmology correspondence \cite{Skenderis:2006jq}.
As ref.~\cite{Skenderis:2007sm} focusses on proper supersymmetry, Wick rotation of the coupling constant
has to be necessarily accompanied by a change of reality conditions on the spinors and, furthermore, 
a Wick rotation of the vector field: the result is a true De Sitter supergravity with its characteristic 
ghost-like vector field,
{\em i.e.\/} the kinetic term for the vector field has the wrong sign (see {\em e.g.\/} \cite{Pilch:1984aw}).
In our construction, however, we do not impose proper supersymmetry and do not change the reality
conditions of the spinors: this avoids the problem of having ghost-like vector fields, implying that in the
limit of vanishing FI-term we recover an ordinary supergravity theory.
%%%%%%%%%%%%%%%%%%%%%%%%%%%%%%%%%%% Ne Toucher Pas  %%%%%%%%%%%%%%%%%%%%%%%%%%%%%%%%%%%%%%%%%%%%%%%%%%%%%%%%%%%%%%%%%%%%%%
\par
The outline of this paper is the following: in section (\ref{sec:DSVector}) we shall set up the fake-Killing spinors equations we are going to solve and 
some information about special geometry and the gauging of isometries in special geometries, needed to understand the set up, are given in appendix
(\ref{appsec:SpecGeom}). In that section we will see that, as we are Wick-rotating the FI-term, the relations between the equations of motion one can 
derive from the integrability equation, are similar to the ones obtained in the supersymmetric case and that the implications as far as the checking 
of equations of motion are concerned are identical: this was to be expected as we are not changing the characteristics of the Killing spinors. 
Similar to the supersymmetric classifications, there are two cases to be considered, namely the ones depending on the norm of the vector one
constructs as a bilinear of the fake-Killing spinors, and the time-like case, {\em i.e.\/} when the norm doesn't vanish, will be treated in section (\ref{sec:VectBil}).
In section (\ref{sec:Null}) we will have a go at the null case, {\em i.e.\/} when the norm of the vector vanishes identically. In that section, we shall ignore the
possible non-Abelian couplings and furthermore will not obtain a complete characterisation; in stead we shall see that the solutions have infinitesimal holonomy
contained in $\mathfrak{sim}(2)$ and discuss the general features such a solution should have. This will be illustrated by two solutions, namely the Nariai
cosmos in the minimal theory in section (\ref{sec:NullSols}) and in section (\ref{sec:Holomorphic}) a general class of solutions with holomorphic scalars
which can be seen as a, back-reacted, intersection of a cosmic string with a Robinson-Bertotti-Nariai solution.
\par
The reader might feel that the generic theories that can be treated in our setting are rather esoteric as their connection with supergravity theories or EYM-$\Lambda$
theories is rather weak: in section (\ref{sec:PotIsNul}) we shall use the well-known fact that in gauged $N=2$ $d=4$ supergravity theories there are choices for the 
FI-terms for which the theory under consideration equals that of the bosonic part of an ungauged supergravity \cite{Cremmer:1984hj}. 
This in fact means that in those cases, our fake-supersymmetric
solutions are nothing more than non-BPS solutions to an ordinary ungauged supergravity. The easiest model in which one can see this happen is the model which
can be obtained by dimensionally reducing minimal $N=1$ $d=5$ supergravity, and we shall discuss some simple solutions to this model and also their uplift to five dimensions.
{}Finally, in section (\ref{sec:Concl}) we shall give our conclusions and a small outlook for related work in higher dimensions, and appendices (\ref{sec:Bil}) and 
(\ref{sec:NullCurv}) contains information about the normalisation of the bilinears and the curvatures for the null-case.
%%%%%%%%%%%%%%%%%%%%%%%%%%%%%%%%%%%%%%%%%%%%%%%%%%%%%%%%%%%%%%%%%%%%%%%%%%%%%%%%%%%%%%%%%%%%%%%%%%%%%%%%%%%%%%%%%
\section{Fake $N=2$ Einstein-Yang-Mills}
\label{sec:DSVector}
%%%%%%
As was said in the introduction, the set-up that we want to consider can be obtained from 
ordinary $N=2$ $d=4$ gauged sugra coupled to vector multiples but no hyper-multiplets, 
by Wick-rotating the Fayet-Iliopoulos term: said differently, we Wick-rotate the constant
tri-holomorphic map $\mathsf{P}^{x}_{\Lambda}\rightarrow i\mathtt{C}_{\Lambda}\delta^{x}_{2}$,
where $\mathtt{C}_{\Lambda}$ are real constant. In supersymmetry the FY-term would gauge
an $U(1)$ in the hyper-multiplets' $SU(2)$, and the effect of the Wick-rotation is that 
we are gauging an $\mathbb{R}$-symmetry through the effective connection $\mathtt{C}_{\Lambda}A^{\Lambda}$ \cite{Behrndt:2003cx}.
\par
The presence of a FI-term is compatible with the gauging of non-Abelian
isometries of the scalar manifold, as long as the action of the gauge group commutes with
the FI-term (see {\em e.g.\/} \cite{Andrianopoli:1996cm}); taking the gauge algebra to 
have structure constants $f_{\Lambda\Sigma}{}^{\Gamma}$, then implies that we must impose
the constraint $f_{\Lambda\Sigma}{}^{\Omega}\mathtt{C}_{\Omega}=0$. 
One result of the introduction of the $\mathtt{C}_{\Lambda}$ is that the dimension of the
possible gauge-algebra is not $\bar{n}=n+1$, $n$ being the number of vector multiplets, 
but rather $n$, as `one` vector field is already used as the connection for the $\mathbb{R}$-symmetry.
\par
The gauging of isometries implies that field-strengths of the physical fields are given by
\begin{equation}
  \label{eq:Deriv1}
  \mathtt{D}Z^{i}\; \equiv\; dZ^{i}\ +\ gA_{\Lambda}^{\Lambda}\ \mathtt{K}^{i} \hspace{.5cm} ,\hspace{.5cm}
  F^{\Lambda} \; \equiv\; dA^{\Lambda} 
        \ +\ \textstyle{g\over 2}\ f_{\Sigma\Gamma}{}^{\Lambda}\ A^{\Sigma}\wedge A^{\Gamma} \; . 
\end{equation}
where $\mathtt{K}_{\Lambda}^{i}$ is the holomorphic part of the Killing vector $\mathtt{K}_{\Lambda}$
(see appendix (\ref{appsec:SpecGeom}) for the minimal information needed 
or refs.~\cite{Andrianopoli:1996cm,Hubscher:2008yz} for a fuller account).
One implication of the above definition is that
$\mathtt{C}_{\Lambda}F^{\Lambda} = d\left[ \mathtt{C}_{\Lambda}A^{\Lambda}\right]$, so that the linear
combination $\mathtt{C}_{\Lambda}A^{\Lambda}$ is indeed an Abelian vector-field.   
\par
As mentioned, we are introducing an $\mathbb{R}$-connection which together
with the existent K\"ahler/$U(1)$-symmetry due to the vector coupling means that we should define the 
covariant derivative on the F-killing spinors as\footnote{
  In the notation that we will follow throughout this article, $\mathbb{D}$ will be the total connection, 
  whereas we will reserve $\mathtt{D}$ for the connection without the $\mathbb{R}$-part and 
  $\mathfrak{D}$ for the K\"ahler-connection, {\em i.e.\/} the connection appearing in ungauged supergravity.
}
\begin{eqnarray}
  \label{eq:16}
  \mathbb{D}_{a}\epsilon_{I} 
       & =& \nabla_{a}\epsilon_{I}
       \, +\, \textstyle{i\over 2}\mathcal{Q}_{a}\epsilon_{I}
       \, +\, \frac{ig}{2}\ A_{a}^{\Lambda}\ \left[\mathtt{P}_{\Lambda}\ +\ i \mathtt{C}_{\Lambda}\right]\epsilon_{I} 
       \nonumber \\ 
    & \equiv& \mathtt{D}_{a}\epsilon_{I} \, -\, \textstyle{g\over 2}\ \mathtt{C}_{\Lambda}A^{\Lambda}_{a}\epsilon_{I}\; ,
\end{eqnarray}
where $\mathtt{P}_{\Lambda}$ is the momentum map corresponding to
an isometry $\mathtt{K}_{\Lambda}$ of the special geometry.
\par
Using the above definitions we can write the fake Killing Spinor Equations as
\begin{eqnarray}
  \label{eq:20}
  \mathbb{D}_{a}\epsilon_{I} & =& -\varepsilon_{IJ}\ \mathcal{T}_{ab}^{+}\gamma^{b}\ \epsilon^{J}
          \; -\; \textstyle{ig\over 4}\ \mathtt{C}_{\Lambda}\mathcal{L}^{\Lambda}\, \gamma_{a}\ \varepsilon_{IJ}\epsilon^{J} \; ,\\
  \label{eq:20b}
  \mathbb{D}_{a}\epsilon^{I} & =& \varepsilon^{IJ}\ \overline{\mathcal{T}^{+}}_{ab}\gamma^{b}\ \epsilon^{J}
          \; -\; \textstyle{ig\over 4}\ \mathtt{C}_{\Lambda}\overline{\mathcal{L}}^{\Lambda}\, \gamma_{a}\ \varepsilon^{IJ}\epsilon_{J} \; ,\\
  & & \nonumber \\
  \label{eq:20c}
  i\slashed{\mathtt{D}}Z^{i}\ \epsilon^{I} & =& 
     -\varepsilon^{IJ}\ \slashed{G}^{i+}\epsilon_{J} 
     \; -\; \mathtt{W}^{i}\ \varepsilon^{IJ}\epsilon_{J} \; ,\\
  %%%
  \label{eq:20d}
  i\slashed{\mathtt{D}}\overline{Z}^{\bar{\imath}}\ \epsilon_{I} & =& 
     -\varepsilon_{IJ}\ \slashed{\overline{G}}^{\bar{\imath}-}\epsilon^{J} 
     \; -\; \overline{\mathtt{W}}^{\bar{\imath}}\ \varepsilon_{IJ}\epsilon^{J} \; ,   
\end{eqnarray}
where for clarity we have given also the rules for $\mathbb{D}_{a}\epsilon^{I}$ and 
${\scriptstyle\slashed{\mathtt{D}}\overline{Z}^{\bar{\imath}}}$ $\!\!\epsilon_{I}$ even though they can be obtained
by complex conjugation from the other 2 rules.
Furthermore, we introduced the abbreviation
\begin{equation}
  \label{eq:Vect1}
  \mathtt{W}^{i} \; =\; -\textstyle{ig\over 2}\ \bar{f}^{i\Lambda}\ 
          \left[ \mathtt{P}_{\Lambda}\ +\ i\mathtt{C}_{\Lambda}\right]
  \hspace{.5cm},\hspace{.5cm}
  \overline{\mathtt{W}^{\bar{\imath}}} \; =\; \overline{\mathtt{W}^{i}} \; ,
\end{equation}
and we used the standard $N=2$ $d=4$ sugra definitions \cite{Andrianopoli:1996cm}
\begin{equation}
  \label{eq:DefFStrength}
  \mathcal{T}^{+}\; \equiv\; 2i\mathcal{L}_{\Lambda}\ F^{\Lambda +} \;\;\; ,\;\;\;
  G^{i+} \; \equiv\; -\bar{f}^{i}_{\Lambda}\ F^{\Lambda\ +} \; .
\end{equation}
\par
The integrability conditions for the above system of equations can easily be calculated and
give rise to
\begin{equation}
  \label{eq:VectInt1}
  \mathcal{B}_{ab}\ \gamma^{b}\epsilon_{I} \; =\; -2i\ \mathcal{L}^{\Lambda}
           \left[
              \slashed{\mathcal{B}}_{\Lambda} \ -\ \mathcal{N}_{\Lambda\Sigma}\slashed{\mathcal{B}}^{\Sigma}
           \right]\ \varepsilon_{IJ}\gamma_{a}\epsilon^{J}\; ,
\end{equation}
where we defined not only the Bianchi identity as $\star \mathcal{B}^{\Lambda} = \mathtt{D}F^{\Lambda}(=0)$
but also
\begin{eqnarray}
  \label{eq:VectEOM1}
  \mathcal{B}_{ab} & =& R_{ab}
     \ +\ 2\mathcal{G}_{i\bar{\jmath}}\mathtt{D}_{(a}Z^{i}\mathtt{D}_{b)}\overline{Z}^{\bar{\jmath}}
     \ +\ 4\mathrm{Im}\left(\mathcal{N}\right)_{\Lambda\Sigma}\left[
               F^{\Lambda}_{ac}F^{\Sigma}_{b}{}^{c} 
               -\textstyle{1\over 4}\eta_{ab}F^{\Lambda}_{cd}F^{\Sigma cd}
           \right]
     \ -\ \textstyle{1\over 2}\eta_{ab}\ \mathtt{V} \; ,\\
   & & \nonumber \\
   \label{eq:VectEOM2}
  \star \mathcal{B}_{\Lambda} & =& \mathtt{D}\left[\ \mathcal{N}_{\Lambda\Sigma}\ F^{\Sigma -}
                  + \overline{\mathcal{N}}_{\Lambda\Sigma}\ F^{\Sigma +}
                 \right] 
                 - \textstyle{g\over 2}\mathrm{Re}\left(
                        \mathtt{K}_{\Lambda \bar{\imath}} \star\mathtt{D}\overline{Z}^{\bar{\imath}}
                     \right) 
           \equiv
             \mathtt{D}F_{\Lambda} - \textstyle{g\over 2}\mathrm{Re}\left(
                        \mathtt{K}_{\Lambda \bar{\imath}} \star\mathtt{D}\overline{Z}^{\bar{\imath}}
                     \right)\; ,\\
  & & \nonumber \\
  \label{eq:VectPot}
  \mathtt{V} & =& \textstyle{g^{2}\over 2}\left[
                       3\mathtt{C}_{\Lambda}\mathtt{C}_{\Sigma}\mathcal{L}^{\Lambda}\overline{\mathcal{L}}^{\Sigma}
                       \, +\, f_{i}^{\Lambda}\ \bar{f}^{i\Sigma}
                           \left(\mathtt{P}+i\mathtt{C}\right)_{\Lambda}
                            \left(\mathtt{P}+i\mathtt{C}\right)_{\Sigma}
                  \right] \; .
\end{eqnarray}
The potential that follows from the integrability condition is not real, and imposing it to be real
implies that we must satisfy the constraint
\begin{equation}
  \label{eq:GaugeConstr}
  0\; =\; \mathrm{Im}\left(\mathcal{N}\right)^{-1|\Lambda\Sigma}\ \mathtt{P}_{\Lambda}\ \mathtt{C}_{\Sigma} \; ,
\end{equation}
which is a gauge-invariant statement. 
{}For our choice of possible non-Abelian gaugings, this constraint is satisfied identically: 
by contracting the last equation in eq.~(\ref{eq:SGK17}) with $f_{i}^{\Sigma}$ and using identities
(\ref{eq:SGImpId}) and (\ref{eq:SGK10}) one can obtain the identity 
\begin{equation}
  \label{eq:GaugeConstr2}
  \mathrm{Im}\left(\mathcal{N}\right)^{-1|\Lambda\Sigma}\ \mathtt{P}_{\Sigma} \; =\; 
      4i\ \mathcal{L}^{\Sigma}\overline{\mathcal{L}}^{\Omega}\ f_{\Sigma\Omega}{}^{\Lambda}\; ,
\end{equation}
which upon contracting with $\mathtt{C}_{\Lambda}$ and using its $G$-invariance gives the desired result.
Therefore the potential $\mathtt{V}$ reads
\begin{eqnarray}
  \label{eq:Potential}
  \mathtt{V}& =& \textstyle{g^{2}\over 2}\left[
                       3\left|\mathtt{C}_{\Lambda}\mathcal{L}^{\Lambda}\right|^{2}
                       \; +\; f_{i}^{\Lambda}\bar{f}^{i\Sigma}\left(\
                              \mathtt{P}_{\Lambda}\mathtt{P}_{\Sigma}\ -\ \mathtt{C}_{\Lambda}\mathtt{C}_{\Sigma}
                        \right)\
                   \right]\; , \\
  & & \nonumber \\
  \label{eq:Potential1}
  & =& \textstyle{g^{2}\over 2}\left[
                       4\left|\mathtt{C}_{\Lambda}\mathcal{L}^{\Lambda}\right|^{2}
                       \; +\; \textstyle{1\over 2}\mathrm{Im}\left(\mathcal{N}\right)^{-1|\Lambda\Sigma}
                        \left(\
                              \mathtt{C}_{\Lambda}\mathtt{C}_{\Sigma}\ -\ \mathtt{P}_{\Lambda}\mathtt{P}_{\Sigma}
                        \right)\
                   \right]\; ,
\end{eqnarray}
which is similar to the supersymmetric result in \cite{Andrianopoli:1996cm}, upon Wick rotating
the Fayet-Iliopoulos term.
Likewise, the above equations of motion can then be obtained from the action
\begin{equation}
  \label{eq:VectAct}
  \int_{4}\sqrt{g}\left[
     R 
    + 2\mathcal{G}_{i\bar{\jmath}}\mathtt{D}_{a}Z^{i}\mathtt{D}^{a}\overline{Z}^{\bar{\jmath}}
    + 2\mathrm{Im}\left(\mathcal{N}\right)_{\Lambda\Sigma}F^{\Lambda}_{ab}F^{\Sigma ab}
    - 2\mathrm{Re}\left(\mathcal{N}\right)_{\Lambda\Sigma}F^{\Lambda}_{ab}\star F^{\Sigma ab}
    - \mathtt{V}
  \right] \; ,
\end{equation}
which as stated in the introduction has correctly normalised kinetic terms.
\par
In sugra the integrability condition for the scalars relates the scalar e.o.m.~with the
Maxwell e.o.m.s, and the same happens here: a straightforward calculation results in 
\begin{equation}
  \label{eq:VectInt2}
  \mathcal{B}^{i}\epsilon_{I} \; =\; -2i\ \bar{f}^{i\Lambda}\left[
                        \slashed{\mathcal{B}}_{\Lambda}
                        \ -\ \mathcal{N}_{\Lambda\Sigma}\slashed{\mathcal{B}}^{\Sigma}
                  \right]\ \varepsilon_{IJ}\epsilon^{J} \; ,
\end{equation}
where we have introduced the equation of motion for the scalars $Z^{i}$ as
\begin{equation}
  \label{eq:VectEOM3}
  \mathcal{B}^{i} \, =\,
      \Box Z^{i} 
     -i\partial^{i}\overline{\mathcal{N}}_{\Lambda\Sigma}F^{\Lambda +}_{ab}F^{\Sigma +\ ab}
     +i\partial^{i}\mathcal{N}_{\Lambda\Sigma}F^{\Lambda -}_{ab}F^{\Sigma -\ ab}
     +\textstyle{1\over 2}\partial^{i}\mathtt{V} \; .
\end{equation}
In conclusion, the integrability conditions for the equations (\ref{eq:20}--\ref{eq:20d}) give
relations between the equations of motion, which, forgetting about the changes in the form of the 
$\mathcal{B}$-tensors, are exactly the same as found in supersymmetry, which is 
hardly surprising. The implication of the relations (\ref{eq:VectInt1}) and (\ref{eq:VectInt2}) 
is then also the same \cite{Gauntlett:2002nw,Bellorin:2005hy}, namely that the independent number of equations of motion
one has to check in order to be sure that a given solution to eqs.~(\ref{eq:20}--\ref{eq:20d})
is also a solution to the equations of motion is greatly reduced.\footnote{
  As we are using the same conventions as ref.~\cite{Meessen:2006tu}, we can copy their arguments as they stand.
}
The minimal set of equations of motion one has to check depends on the norm of the vector
bilinear $V_{a}=i\overline{\epsilon}^{I}\gamma_{a}\epsilon_{I}$: 
if the norm $V_{a}V^{a}$ is positive, referred to as the time-like case, we only need to solve
the time-like direction of the Bianchi identity, {\em i.e.\/} $\imath_{V}\star\mathcal{B}^{\Lambda}=0$ 
and the Maxwell/YM equations, {\em i.e.\/} $\imath_{V}\star\mathcal{B}_{\Lambda}=0$. This case will 
be considered in section (\ref{sec:VectBil}).
\par
If the norm of the bilinear is null, {\em i.e.\/} $V_{a}V^{a}=0$, then a convenient set of e.o.m.s is given by
$N^{a}N^{b}\mathcal{B}_{ab}=0$, $N^{a}\mathcal{B}_{\Lambda a}=0$ and $N^{a}\mathcal{B}_{a}^{\Lambda}$,
where $N$ is a vector normalised by $V^{a}N_{a}=1$: 
this case will be considered in section (\ref{sec:Null}).
%
%%%%%%%%%%%%%%%%%%%%%%%%%%%%%%%%%%%%%%%%%%%%%%%%%%%%%%%%%%%%%%%%%%%%%%%%%%%%%%%%%%%%%%%%%%%%%%%%%%%%%%%%%%%%%%%%%
\section{Analysis of the Time-like case}
\label{sec:VectBil}
%%%%%%
In this section we shall consider the time-like case and the strategy to be followed is the usual one:
we analyse the differential constraints on the bilinears constructed out of the spinors $\epsilon_{I}$
defined in appendix (\ref{sec:Bil}) coming from the fKSEs (\ref{eq:20}--\ref{eq:20d}), trying to solve these
constraints as general as possible in as little unknowns as possible. After the constraints have been solved, we shall, following the comments
made above, impose the Bianchi identity and the gauge-field equations of motion and to see what conditions they impose.
After these steps we will be left with a minimal set of functions, structures and conditions they have to satisfy in order
to construct fake-supersymmetric solutions: for the solutions to be constructed in this case, the algorithm will be outlined in
section (\ref{sec:CosmMon}).
\par
Let us start by discussing the differential constraints on the bilinears: 
using eq.~(\ref{eq:20}) and the definitions of the bilinears in appendix (\ref{sec:Bil}), we can calculate
\begin{eqnarray}
  \label{eq:VId1}
  \mathbb{D}X & =&  \textstyle{g\over 4}\ \mathtt{C}_{\Lambda}\mathcal{L}^{\Lambda}\ V
           \; +\; i\ \imath_{V}\mathcal{T}^{+}\; ,\\
    & & \nonumber\\
  \label{eq:VId2}
  \mathbb{D}_{a}V_{b} & =&  g|X|^{2}\ \mathtt{C}_{\Lambda}\mathcal{R}^{\Lambda}\ \eta_{ab}
      \ +\ 4\mathrm{Im}\left(
                \overline{X}\ \mathcal{T}^{+}_{ab}
            \right)\; ,\\
   & & \nonumber \\
  \label{eq:VId3}
  \mathbb{D}V^{x} & =& \textstyle{g\over 2}\mathtt{C}_{\Lambda}\mathcal{R}^{\Lambda}\ V\wedge V^{x}
     \ +\ \textstyle{g\over 2}\mathtt{C}_{\Lambda}\mathcal{I}^{\Lambda}\
            \star\left[ V\wedge V^{x}\right] \; ,
\end{eqnarray}
where following ref.~\cite{Meessen:2006tu} we have introduced the real symplectic
sections of K\"ahler weight zero,
\begin{equation}
  \label{eq:DefRandI}
  \mathcal{R} \; =\; \mathrm{Re}\left(\mathcal{V}/X\right) \;\;\; ,\;\;\;
  \mathcal{I} \; =\; \mathrm{Im}\left(\mathcal{V}/X\right) \;\longrightarrow\;\;
  \frac{1}{2|X|^{2}}\ =\ \langle\mathcal{R}|\mathcal{I}\rangle \; .
\end{equation}
In the ungauged theory, as will also be the case here, the $2\bar{n}$ real functions $\mathcal{I}$ play a fundamental r\^ole in the construction
of BPS solutions and the $2\bar{n}$ real functions $\mathcal{R}$ depend on $\mathcal{I}$: finding, given a 
Special Geometric model, the explicit $\mathcal{I}$-dependence of $\mathcal{R}$ is known as the 
{\em stabilisation equation}, and for many models solutions to it are known.
\par
A first difference with supersymmetric case lies in the character of the bilinear $V$: in that case it is always a Killing vector,
which as one can see from eq.~(\ref{eq:VId2}) will not be the case here. We can still use it to introduce a time-like coordinate $\tau$
by choosing  an adapted coordinate system through $V^{a}\partial_{a}=\sqrt{2}\partial_{\tau}$, but now the components of the metric
will depend explicitly on $\tau$, as was to be expected from for instance the Kastor {\&} Traschen solution \cite{Kastor:1993mj}.
\par
As the $V^{x}$ contain the information about the metric on the base-space, it is important to deduce its 
behaviour under translations along $V$; in order to investigate we calculate
\begin{equation}
  \label{eq:LieV}
  \pounds_{V}V^{x} \; =\; \imath_{V}dV^{x} +d\left(\imath_{V}V^{x}\right)
               \; =\; g \mathtt{C}_{\Lambda}\imath_{V}A^{\Lambda}\ V^{x} 
                      \ +\ 2g|X|^{2}\mathtt{C}_{\Lambda}\mathcal{R}^{\Lambda}\ V^{x} \; .
\end{equation}
This implies that by choosing the gauge-fixing
\begin{equation}
  \label{eq:VId4}
  \imath_{V}A^{\Lambda} \; =\; -2|X|^{2}\ \mathcal{R}^{\Lambda} \; ,
\end{equation}
we find that $\pounds_{V}V^{x}=0$. We would like to point out that the above gauge-fixing
is the actual result one obtains when considering time-like supersymmetric solutions in 
$N=2$ $d=4$ supergravity theories \cite{Meessen:2006tu,Hubscher:2008yz}.
\par
The above result has some nice implications, the first of which is derived by contracting eq.~(\ref{eq:VId2})
with $V^{a}V^{b}$, namely
\begin{equation}\label{eq:Res1}
  \langle\nabla_{V}\mathcal{R}|\mathcal{I}\rangle + \langle\mathcal{R}|\nabla_{V}\mathcal{I}\rangle 
  \; =\; \nabla_{V}\frac{1}{2|X|^{2}} 
  \; =\; g\mathtt{C}_{\Lambda}\ \mathcal{R}^{\Lambda} \; .
\end{equation}
We can rewrite the above equation to a nicer form by observing that 
\begin{eqnarray}
  \langle \mathcal{V}/X| d\left(\mathcal{V}/X\right)\rangle & =& 
          X^{-2}\ \langle\mathcal{V}|\mathfrak{D}\mathcal{V}\rangle
          \ -\ X^{-3}\mathfrak{D}X\; \langle\mathcal{V}|\mathcal{V}\rangle \; =\; 0\nonumber \\
     & =& \langle\mathcal{R}|d\mathcal{R}\rangle \ -\ \langle\mathcal{I}|d\mathcal{I}\rangle
          \, +\, i\langle\mathcal{R}|d\mathcal{I}\rangle
          \, +\, i\langle\mathcal{I}|d\mathcal{R}\rangle \; ,
\end{eqnarray}
which seeing the reality properties of the above expression implies\footnote{
  These expressions were derived in ref.~\cite{Bellorin:2006xr} starting from a prepotential
  and using the obvious homogeneity of the symplectic section $\mathcal{R}$. The derivation
  presented here is far less involved and also holds in situations where no prepotential exists.
}
\begin{eqnarray}
  \label{eq:StabId1}
  \langle\ d\mathcal{R}\ |\ \mathcal{I}\ \rangle & =& \langle\ \mathcal{R}\ |\ d\mathcal{I}\ \rangle \; ,\\
  \label{eq:StabId2}
  \langle\ \mathcal{R}\ |\ d\mathcal{R}\ \rangle & =& \langle\ \mathcal{I}\ |\ d\mathcal{I}\ \rangle \; .
\end{eqnarray}
If we then introduce the real symplectic section $\mathtt{C}^{T}=(0,\mathtt{C}_{\Lambda})$, we can rewrite
eq.~(\ref{eq:Res1}) in the simple and suggestive form
\begin{equation}
  \label{eq:VId5a}
  0\; =\; \langle\ \mathcal{R}\ |\ \nabla_{V}\mathcal{I}\ +\ \textstyle{g\over 2}\mathtt{C}\ \rangle \; .
\end{equation}
The above equation could also have been obtained from the contraction of eq.~(\ref{eq:VId1}) with $V$, {\em i.e.\/}
\begin{equation}
  \label{eq:VId5b}
  \textstyle{\frac{1}{\overline{X}}}\ \mathtt{D}_{V}\textstyle{\frac{1}{X}}
    \; =\; -g\langle\mathcal{R}|\mathtt{C}\rangle
            \, +\, ig\langle\mathcal{I}|\mathtt{C}\rangle \; ,
\end{equation}
and taking its real part.
By taking the imaginary part and using the identity
\begin{equation}
  \mathrm{Im}\left(\ \textstyle{\frac{1}{\overline{X}}}\ \mathtt{D}\textstyle{\frac{1}{X}}\ \right)
  \, =\; -2\langle\mathcal{I}|\mathtt{D}\mathcal{I}\rangle \; ,
\end{equation}
we find that apart form eq.~(\ref{eq:VId5a}), we also must have
\begin{equation}
  \label{eq:VId5c}
  0\; =\; \langle\ \mathcal{I}\ |\ \nabla_{V}\mathcal{I}\ +\ \textstyle{g\over 2}\mathtt{C}\ \rangle \; .
\end{equation}
\par
By now, there are strong hints that the derivative of the symplectic section $\mathcal{I}$ in the 
direction $V$ should be constant and, in fact, the information needed to close the case is hidden
in eqs.~(\ref{eq:20c}) and (\ref{eq:20d}).
{}From the contraction of (\ref{eq:20c}) with $\bar{\epsilon}^{K}\gamma_{a}\varepsilon_{KI}$ we find
\begin{equation}
  \label{eq:VId6}
  2\overline{X}\ \mathtt{D}Z^{i} \; =\; 4\ \imath_{V}G^{i+} \ -\ \mathtt{W}^{i}\ V\; ,
\end{equation}
which upon contraction with $V$ leads to 
\begin{equation}
  \label{eq:VId7}
  \mathtt{D}_{V}Z^{i} \; =\; -2\ X\ \mathtt{W}^{i} \; .
\end{equation}
Using the gauge-fixing (\ref{eq:VId4}), the identity 
$\bar{f}^{\Lambda i}\mathtt{P}_{\Lambda}=i\overline{\mathcal{L}}^{\Lambda}\mathtt{K}_{\Lambda}^{i}$
and the fact that for our choice of possible non-Abelian gauge groups we have 
$\mathcal{L}^{\Lambda}\mathtt{K}_{\Lambda}^{i}=0$,
we see that the above equation is converted to 
\begin{equation}
  \label{eq:VId7a}
  \nabla_{V}Z^{i} \; =\; -g\ X\ \bar{f}^{\Lambda i}\ \mathtt{C}_{\Lambda} \; .
\end{equation}
Using then the special geometry identity
$\langle\mathcal{U}_{i}|\overline{\mathcal{U}}_{\bar{\jmath}}\rangle = i\mathcal{G}_{i\bar{\jmath}}$,
we can rewrite the above equation to
\begin{equation}
  \langle \nabla_{V}\mathcal{I}+g\mathtt{C}\, |\ \overline{\mathcal{U}}_{\bar{\jmath}}\rangle
    \; =\; i\langle\ \nabla_{V}\mathcal{R}|\overline{\mathcal{U}}_{\bar{\jmath}}\rangle \; ,
\end{equation}
which can be manipulated by using the special geometry properties and a renewed call to 
eq.~(\ref{eq:VId7}) to give
\begin{equation}
  \label{eq:Vad6}
  \langle\ \nabla_{V}\mathcal{I}\ +\ \textstyle{g\over 2}\mathtt{C}\, |\ \overline{\mathcal{U}}_{\bar{\jmath}}\rangle
   \; =\; 0 \; .
\end{equation}
The above equation plus eqs.~(\ref{eq:VId5a}) and (\ref{eq:VId5c}) together with the completeness relation
from special geometry, eq.~(\ref{eq:SGSymplProj}), then implies 
\begin{equation}
  \label{eq:LaPolla}
  \nabla_{V}\mathcal{I} \; =\; -\textstyle{g\over 2}\ \mathtt{C} \; ,
\end{equation}
which implies that the $\tau$-dependence of the functions $\mathcal{I}$ is at most linear, and in fact only
half of them, namely the $\mathcal{I}_{\Lambda}$.
\par
At this point it is necessary to introduce a complete coordinate system ($\tau, y^{m}$), which we will take to be adapted to
$V$ and compatible with the Fierz identities in appendix (\ref{sec:Bil}), {\em i.e.\/}
\begin{equation}
  \label{eq:Coord}
  \begin{array}{lclclcl}
    V^{a}\partial_{a} & =& \sqrt{2}\partial_{\tau} &\hspace{.3cm},\hspace{.3cm}&
    V & =& 2\sqrt{2}|X|^{2}\left( d\tau \ +\ \omega\right) \\
    & & & & & & \\
    V^{xa}\partial_{a} & =& -2\sqrt{2}|X|^{2}V^{xm}\ \left(\partial_{m}-\omega_{m}\partial_{\tau}\right)
    & ,& 
    V^{x} & =& \sqrt{2}\ V^{x}_{m}\ dy^{m} \; ,
  \end{array}
\end{equation}
where $\omega =\omega_{m}dy^{m}$ is a possibly $\tau$-dependent 1-form
and we introduced $V^{xm}$ by $V^{xm}V^{y}_{m}=\delta^{xy}$;
as the $V^{x}_{m}$ act as a Dreibein on a Riemannian space, the $x$-indices can be raised and lowered
with $\delta^{xy}$, so that we won't distinguish between co- and contravariant $x$-indices.
\par
Putting the Vierbein together with the Fierz identity (\ref{eq:Bil4}) we find that the metric is takes on the
conforma-stationary form
\begin{equation}
  \label{eq:37}
  ds^{2} \; =\; 2|X|^{2}\left( d\tau \ +\ \omega\right)^{2}
      \ -\ \textstyle{1\over 2|X|^{2}}\ h_{mn} dy^{m}dy^{n} \; ,
\end{equation}
where $h_{mn}=V^{x}_{m}V^{x}_{n}$ is the metric on the 3-dimensional base-space.
\par
W.r.t.~our choice of coordinates we have that $\pounds_{V}V^{x}=0$ equals $\partial_{\tau}V^{x}_{m}=0$;
the $V^{x}$ are of course also constrained by eq.~(\ref{eq:VId3}), which in the chosen coordinate system
and using the decomposition
\begin{equation}
  \label{eq:Dec1}
  A^{\Lambda} \; =\; -\textstyle{1\over 2}\mathcal{R}^{\Lambda}\ V \ +\ \tilde{A}^{\Lambda}_{m}\ dy^{m}
             \ \equiv\ -\textstyle{1\over 2}\mathcal{R}^{\Lambda}\ V \ +\ \tilde{A}^{\Lambda} 
  \;\ \longrightarrow\;\ F^{\Lambda} \ =\ -\textstyle{1\over 2}\mathtt{D}\left[\mathcal{R}^{\Lambda}V\right]
                                     \ +\ \tilde{F}^{\Lambda} \; ,
\end{equation}
reads
\begin{equation}
  \label{eq:ResVx}
  dV^{x} \; =\; g\mathtt{C}_{\Lambda}\ \tilde{A}^{\Lambda} \wedge V^{x}
         \ +\ \textstyle{g\over 4}\ \mathtt{C}_{\Lambda}\mathcal{I}^{\Lambda}\ 
                \varepsilon^{xyz}\ V^{y}\wedge V^{z}\; .
\end{equation}
A first remark to be made is that for consistency we must have 
$\mathtt{C}_{\Lambda}\partial_{\tau}\tilde{A}^{\Lambda}_{x}=0$.
Further, we could use the residual gauge freedom 
$\mathtt{C}_{\Lambda}\tilde{A}^{\Lambda}\rightarrow \mathtt{C}_{\Lambda}\tilde{A}^{\Lambda} + d\phi (y)$, 
$V^{x}\rightarrow e^{g\phi}V^{x}$ to take $\mathtt{C}_{\Lambda}\mathcal{I}^{\Lambda}$ to be constant, a possibility we will not use.
And lastly, the integrability condition $d^{2}V^{x}=0$ implies
\begin{equation}
  \label{eq:ResVxInt}
  0\; =\; \textstyle{g\over 4} \left[\
             \varepsilon^{xyz}\ \mathtt{C}_{\Lambda}\tilde{F}^{\Lambda}_{yz} 
              \ +\ \sqrt{2}\ V_{x}^{m} \tilde{\mathbb{D}}_{m}\ \mathtt{C}_{\Lambda}\mathcal{I}^{\Lambda}
          \right] \; ,
\end{equation}
where we have introduced $\tilde{F}^{\Lambda}_{xy}\equiv V_{x}^{m}V_{y}^{n}\tilde{F}^{\Lambda}_{mn}$
and 
\begin{equation}\label{eq:DefTildeD}
    \tilde{\mathbb{D}}_{m}\mathcal{I}
    \; =\; \partial_{m} \mathcal{I}
                   \ +\ g\mathtt{C}_{\Lambda}\tilde{A}^{\Lambda}_{m}\ \mathcal{I}
                   \ +\ g\tilde{A}^{\Lambda}_{m}\ S_{\Lambda}\mathcal{I} \;\;\; ;\;\; 
    \tilde{\mathbb{D}}_{x} \ \equiv\ V_{x}^{m}\ \tilde{\mathbb{D}}_{m} \; .
\end{equation}
\par
The system (\ref{eq:ResVx}) was analysed by Gauduchon {\&} Tod in ref.~\cite{Gauduchon:1998}, as it appeared in the discussion
of 4-dimensional hyper-hermitian Riemannian metrics admitting a tri-holomorphic Killing vector. A first implication is that the
geometry of the base-space belongs to a subclass of 3-dimensional Einstein-Weyl spaces, called hyper-CR or Gauduchon-Tod spaces: one of the 
extra constraints to be imposed on the EW-spaces is nothing more than the integrability condition (\ref{eq:ResVxInt})
which is called the {\em generalised Abelian monopole equation}.
As we will see later on, and can be expected from the similar discussion in \cite{Hubscher:2008yz}, the equations determining the 
seed function $\mathcal{I}^{\Lambda}$, will be generalised non-Abelian monopole equation or, said differently, the straightforward
generalisation of the standard Bogomol'nyi equation on $\mathbb{R}^{3}$ to GT-spaces; eq.~(\ref{eq:ResVxInt}) is of course implied by these upon contraction
with $\mathtt{C}_{\Lambda}$.
\par
In ref.~\cite{Behrndt:2003cx}, Behrndt and Cveti\v{c} realised that their 5-dimensional cosmological solutions could be dimensionally
reduced to 4-dimensional ones, which raises the question of what solution found by Grover {\em et al.\/} \cite{Grover:2008jr} can be reduced to 
solutions we are going ot find. As in this case we are dealing with a map between the 5-dimensional time-like case and the 
4-dimensional time-like case, the dimensional reduction has to be over the 4-dimensional base-space, which was found to 
be hyperK\"ahler-torsion \cite{Grover:2008jr}. The key to identifying the subclass of 5-dimensional solutions that can be reduced to
ours, then lies in a further result of Gauduchon {\&} Tod (see remark 2 in ref.~\cite{Gauduchon:1998}), which states that the solutions
to eqs.~(\ref{eq:ResVx}) and (\ref{eq:ResVxInt}) are obtained by the reduction of a conformal hyper-K\"ahler space along a tri-holomorphic
Killing vector. In fact, as is nicely discussed in \cite[sec.~(3.2)]{Grover:2008jr}, these spaces are particular instances of HKT-spaces.
This inheritance of geometrical structures also ocurrs in ordinary supergravity theories in $6$, $5$ and $4$ dimensions and it is reasonable
to suppose that this also holds for fake/Wick-rotated supergravities.
As a final comment, let us mention that the 3-dimensional Killing spinor equation on a GT-manifold
allows non-trivial solutions \cite{Buchholz:2000}.
\par
Before turning to the equations of motion, we deduce the following equation for $\omega$ from the
anti-symmetrised version of eq.~(\ref{eq:VId2}) and the explicit coordinate expression in (\ref{eq:Coord}).
As the reader will observe, this calculation needs the explicit form for the 2-form $\mathcal{T}^{+}$,
which can be obtained from eq.~(\ref{eq:VId1}) and the rule that an general imaginary self-dual 2-form
$B^{+}$ is determined by its contraction with $V$ by means of (See refs.~\cite{Caldarelli:2003pb} for more detail)
\begin{equation}
  \label{eq:Tplus}
  B^{+} \; =\; \frac{1}{4|X|^{2}}\left(\
                 V\wedge\ \imath_{V}B^{+} 
                 \ +\ i\star\left[\
                           V\wedge\ \imath_{V}B^{+}   
                       \right]  
               \ \right)\; .
\end{equation}
The result reads
\begin{equation}
  \label{eq:Rot}
  d\omega \ +\ g\mathtt{C}_{\Lambda}\tilde{A}^{\Lambda}\wedge (d\tau +\omega ) \; =\; 
  \sqrt{2}\ \star\left[ V\wedge\ \langle\mathcal{I}|\ \mathtt{D}\mathcal{I}\rangle\right] \; .
\end{equation}
Contracting the above equation with $V$ we find that 
\begin{equation}
  \label{eq:OmegaV}
   \pounds_{V}\omega \; =\; g\sqrt{2}\mathtt{C}_{\Lambda}\tilde{A}^{\Lambda} \hspace{.5cm}\longrightarrow\hspace{.5cm}
   \omega \; =\; g\mathtt{C}_{\Lambda}\tilde{A}^{\Lambda}\ \tau \; +\; \varpi \; ,
\end{equation}
where $\varpi = \varpi_{m}dy^{m}$ is $\tau$-independent.
Substituting the above result into eq.~(\ref{eq:Rot}) and evaluating its r.h.s., we obtain
\begin{equation}
  \label{eq:Rot2}
  d\varpi \ +\ g\mathtt{C}_{\Lambda}\tilde{A}^{\Lambda}\wedge \varpi +g\mathtt{C}_{\Lambda}\tilde{F}^{\Lambda}\ \tau\; =\; 
      \textstyle{1\over 2}\ \langle\mathcal{I}|\
           \tilde{\mathtt{D}}_{m}\mathcal{I}\ -\ \omega_{m}\partial_{\tau}\mathcal{I}\rangle
      \ V^{xm}\varepsilon^{xyz}\ V^{y}\wedge V^{z}\; .
\end{equation}
There is a possible inconsistency in this equation due to the possible $\tau$-dependence in the above equation;
as the equation is at most linear in $\tau$, we can investigate the possible inconsistency by taking
the $\tau$-derivative, only to find eq.~(\ref{eq:ResVxInt}). The equation determining $\varpi$ is then found
by splitting off the $\tau$-dependent part and reads
\begin{equation}
  \label{eq:Rot3}
  \tilde{\mathbb{D}}\ \varpi \; =\; 
     \textstyle{1\over 2}\ \varepsilon^{xyz}\ \langle \tilde{\mathcal{I}}\ |\ 
            \tilde{\mathtt{D}}_{x}\tilde{\mathcal{I}} - \varpi_{x}\partial_{\tau}\mathcal{I}\rangle\
            V^{y}\wedge V^{z} \; ,
\end{equation}
where we introduced $\tilde{\mathcal{I}}=\mathcal{I}(\tau =0)$.
\par
The symplectic field strength $F^{T}=(F^{\Lambda},F_{\Lambda})$ then easily be deduced to give the standard
supersymmetric result
\begin{eqnarray}
  \label{eq:Fsympl}
  F & =& -\textstyle{1\over 2}\ \mathtt{D}\left( \mathcal{R}\ V\right)
    \; -\; \textstyle{1\over 2}\ \star\left[
              V\wedge\ \mathbb{D}\mathcal{I}
           \right] \nonumber \\ 
    & =& -\textstyle{1\over 2}\ \mathtt{D}\left( \mathcal{R}\ V\right) 
    \; -\; \frac{\sqrt{2}}{8}\ \varepsilon^{xyz}\ V_{x}^{m}\left[\
                    \tilde{\mathbb{D}}_{m}\mathcal{I}\ -\ \omega_{m}\partial_{\tau}\mathcal{I}
                \ \right]\ V^{y}\wedge V^{z} \; ,
\end{eqnarray}
which agrees completely with the imposed gauge-fixing (\ref{eq:VId4}).
\par
At this point we would like to treat the Bianchi identity $\mathtt{D}F^{\Lambda}=0$, as it was treated in ref.~\cite{Hubscher:2008yz},
namely as leading to a Bogomol'nyi equation determining the pair $(\tilde{A}^{\Lambda},\mathcal{I}^{\Lambda})$; this
approach boils down to stating that since we are given the potential in eq.~(\ref{eq:Dec1}), the Bianchi
identity is solved identically. This does, however, not mean that any given $\tilde{A}^{\Lambda}$ leads
to a field strength with the form prescribed by fake-supersymmetry in eq.~(\ref{eq:Fsympl}).
If we then impose that a given $\tilde{A}^{\Lambda}$ leads to a field-strength with the prescribed form implies imposing
the equation
\begin{equation}
  \label{eq:Bogo}
  \tilde{F}^{\Lambda}_{xy} \; =\; -\frac{1}{\sqrt{2}}\ \varepsilon^{xyz}\ \tilde{\mathbb{D}}_{z}\mathcal{I}^{\Lambda} \; ,
\end{equation}
which due to eq.~(\ref{eq:LaPolla}) is manifestly $\tau$-independent.
This equation is the generalisation of the standard Bogomol'nyi equation on $\mathbb{R}^{3}$ to 
a 3-dimensional Gauduchon-Tod space.
Clearly, the above equation implies the constraint (\ref{eq:ResVxInt}) upon contraction with $\mathtt{C}_{\Lambda}$.
\par
In order to show that the time-like solutions to the fKSEs we characterised are indeed solutions to the equations of 
motion, we need to impose the Maxwell-Yang-Mills equations of motion, {\em i.e.\/} eq.~(\ref{eq:VectEOM2}). 
This equation consists of 2 parts, namely one in the time-direction, {\em e.g.\/} $\mathcal{B}_{\Lambda}^{t}$,
and one in the space-like directions, $\mathcal{B}_{\Lambda}^{x}$.
A tedious but straightforward calculation shows that $\mathcal{B}_{\Lambda}^{t}=0$
identically, in full concordance with the discussion in section (\ref{sec:DSVector});
the equations of motion in the $x$-direction, however, do not vanish identically. In stead, they impose the
condition
\begin{equation}
  \label{eq:MYM1}
  \left(\tilde{\mathbb{D}}_{x}\ -\ \omega_{x}\partial_{\tau}\right)^{2}\mathcal{I}_{\Lambda} \; =\; 
     \textstyle{g^{2}\over 2}\ f_{\Lambda (\Omega}{}^{\Gamma}f_{\Delta )\Gamma}{}^{\Sigma}
          \mathcal{I}^{\Omega}\mathcal{I}^{\Delta}\ \mathcal{I}_{\Sigma}
     \ -\ \textstyle{g^{2}\over 2}\ f_{\Lambda\Omega}{}^{\Sigma}\mathcal{I}^{\Omega}\mathcal{I}_{\Sigma}\
              \mathtt{C}_{\Gamma}\mathcal{I}^{\Gamma}\; ,
\end{equation}
which in the limit $\mathtt{C}\rightarrow 0$ coincides with the result obtained in ref.~\cite{Hubscher:2008yz}.
A simplification of the above equation can be obtained by observing that, due to eqs.~(\ref{eq:DefTildeD}) and 
(\ref{eq:OmegaV}), 
\begin{equation}
 \partial_{\tau}\left(\tilde{\mathbb{D}}_{m}\mathcal{I}_{\Lambda}\ -\ \omega_{m}\partial_{\tau}\mathcal{I}_{\Lambda}\right)
    \; =\; \partial_{\tau}\partial_{m}\mathcal{I}_{\Lambda} \; =\; 0\; .
\end{equation}
Using the above identity and using the fact that $\mathcal{I}_{\Lambda}$ is linear in $\tau$, we can rewrite
eq.~(\ref{eq:MYM1}) as
\begin{equation}
  \label{eq:MYM2}
  \tilde{\mathbb{D}}_{x}^{2}\ \tilde{\mathcal{I}}_{\Lambda} 
  \ -\ \left(\tilde{\mathbb{D}}_{x}\varpi_{x}\right)\ \partial_{\tau}\mathcal{I}_{\Lambda}
  \ =\ 
  \textstyle{g^{2}\over 2}\ f_{\Lambda (\Omega}{}^{\Gamma}f_{\Delta )\Gamma}{}^{\Sigma}
    \mathcal{I}^{\Omega}\mathcal{I}^{\Delta}\ \tilde{\mathcal{I}}_{\Sigma}
    \ -\ \textstyle{g^{2}\over 2}\ f_{\Lambda\Omega}{}^{\Sigma}\mathcal{I}^{\Omega}\tilde{\mathcal{I}}_{\Sigma}\
    \mathtt{C}_{\Gamma}\mathcal{I}^{\Gamma}\; ,
\end{equation}
which is a $\tau$-independent equation! 
%\par
% In the supersymmetric case, {\em i.e.\/} when $\mathtt{C}_{\Lambda}=0$,
% one can use the freedom $\tau\rightarrow\tau - f(y)$ to impose the condition $\partial_{x}\varpi_{x}=0$.
% In the case at hand, said transformation induces the shift freedom 
% $\varpi\rightarrow\varpi -\tilde{\mathbb{D}}f$, and then the question is whether one can always
% find an $f$ such that $\tilde{\mathbb{D}}_{x}^{2}f=\tilde{\mathbb{D}}_{x}\varpi_{x}$ in order to
% be able to put $\tilde{\mathbb{D}}_{x}\varpi_{x}=0$. Perhaps the mathematicians have something
% to say about this, and if not, I'll just forget about it.
%%%%%%%%%%%%%%%%%%%%%%%%%%%%%%%%%%%%%%%%%%%%%%%%%%%%%%%%%%%%%%%%%%%%%%%%%%%%%%%%%%%%%%%%%%%%%%%%%%%%%%%%%%%%%%%%%
\subsection{Recapitulation and some comments}
\label{sec:CosmMon}
%%%%%%%
Let us, before making some comments on the generic behaviour of the solutions, spell out the way how to construct solutions
using the results obtained in the foregoing section: the first step is to decide which model to consider, {\em i.e.\/}
one has to specify what special geometric manifold is to be used, 
what non-Abelian groups can and will be gauged, and furthermore the 
constants $\mathtt{C}_{\Lambda}$. 
Given the model, we must then decide what 3-dimensional hyperCR/GT space we are going to use to describe the geometry
of the 3-dimensional base-space; this is equivalent to finding the triple $(V^{x} , \mathtt{C}_{\Lambda}\tilde{A}^{\Lambda},\mathtt{C}_{\Lambda}\mathcal{I}^{\Lambda})$
solving eq.~(\ref{eq:ResVx}).
This decision, then, allows us in principle to solve the Bogomol'nyi equation (\ref{eq:Bogo}) as to determine $(\tilde{A}^{\Lambda},\mathcal{I}^{\Lambda})$.
\par
The next step would be to determine the $\tau$-independent part of the seed functions $\mathcal{I}_{\Lambda}$, remember that their $\tau$-dependence is
fixed by eq.~(\ref{eq:LaPolla}), using equation (\ref{eq:MYM2}). As this equation contains not only the $\mathcal{I}_{\Lambda}$ but also $\varpi$,
we are forced to determine both objects and make sure that eq.~(\ref{eq:Rot3}) is satisfied.
Having gone through the above steps, all that needs to be done is to determine the field-strengths by means of eq.~(\ref{eq:Fsympl}), write down the 
physical scalars $Z^{i}=\mathcal{L}^{i}/\mathcal{L}^{0}$ and the metric by determining the stationarity 1-form $\omega$ by eq.~(\ref{eq:OmegaV})
and the metrical factor $|X|^{2}$ through eq.~(\ref{eq:DefRandI}).
As usual the explicit construction of the last fields goes through the solution of the {\em stabilisation equation} which determines the symplectic
section $\mathcal{R}$ in terms of the seed functions $\mathcal{I}$; for many models, solutions to the stabilisation equations are known. 
\par
As mentioned in the previous section, the 3-dimensional Einstein-Weyl spaces that occur as the geometry of the base-space,
can be obtained by reduction of hyper-K\"ahler spaces
along a tri-holomorphic conformal Killing vector (see  ref.~\cite[sec.~(3.2)]{Grover:2008jr} for detailed information),
which would put us in a position
to discuss the solutions to the Bogomol'nyi equation (\ref{eq:Bogo}). 
However, knowing only explicit solutions to the non-Abelian Bogomol'nyi equation on $\mathbb{R}^{3}$,\footnote{
 Observe that this is a purely non-Abelian restriction as hyperCR/GT-metrics are known, see {\em e.g.\/} \cite{Dunajski:1999qs}
}
means that for the moment the only non-trivial
non-Abelian solutions we can build are the ones that follow from the supersymmetric ones satisfying 
$\mathtt{C}_{\Lambda}\mathcal{I}^{\Lambda}=0$, 
which implies that $\mathtt{C}_{\Lambda}\tilde{A}^{\Lambda}$ is gauge trivial so that the base-space is 
$\mathbb{R}^{3}$, by substituting 
$\mathcal{I}_{\Lambda}\rightarrow \mathcal{I}_{\Lambda}-g\mathtt{C}_{\Lambda}\tau /(2\sqrt{2})$.
 \par
As the base-space is $\mathbb{R}^{3}$, the equations determining the $\tau$-independent part of the $\mathcal{I}$, 
eqs.~(\ref{eq:Bogo}) and (\ref{eq:MYM2}), reduce to the ones for $N=2$ EYM deduced in ref.~\cite{Hubscher:2008yz}: 
indeed the only difference lies in the divergence of $\varpi$ occurring in eq.~(\ref{eq:MYM2}), 
and in the $\mathbb{R}^{3}$-case there is no obstruction to choosing it to vanish from the onset.
\par
At this point, then, the construction of fake-supersymmetric solutions boils down to the substitution
principle put forward by Behrndt {\&} Cveti\v{c} in ref.~\cite{Behrndt:2003cx}: given a supersymmetric
solution to $N=2$ $d=4$ EYM supergravity, Abelian \cite{Behrndt:1997ny}
or non-Abelian \cite{Hubscher:2008yz,Meessen:2008kb},
substitute
$\mathcal{I}_{\Lambda}\rightarrow \mathcal{I}_{\Lambda}-\mathtt{C}_{\Lambda}/(2\sqrt{2})\ \tau$ and impose
the restriction $\mathtt{C}_{\Lambda}\mathcal{I}^{\Lambda}=0$. Of course, when dealing with non-Abelian
gauge groups, not all choices for $\mathtt{C}_{\Lambda}$ are possible, as one must respect the constraint
$f_{\Lambda\Sigma}{}^{\Gamma}\mathtt{C}_{\Gamma}=0$.
\par
The first observation is that generically the asymptotic form of the solution is not De Sitter but rather
Kasner, {\em i.e.\/} the $\tau$-expansion of the base-space is power-like, making the definition of asymptotic
mass even more cumbersome than in the De Sitter case.\footnote{
  Let us in passing point out that in the resulting Kasner spaces there is a time-like conformal
  isometry of the kind used in ref.~\cite{Kastor:2002fu} to define a conformal energy.
}
The second observation is that the metric has a curvature singularity at those events/points
for which $|X|^{-2}=0$, which may be located outside our chosen coordinate system. This, however,
raises the question of the possibility having an horizon, or said differently, how to
decide in a practical manner when our solution describes a black hole. 
Observe that in the original Kastor {\&} Traschen
solution for one single black hole, this question is readily resolved by changing coordinates as to
obtain the time-independent, spherically symmetric extreme RNDS black hole, for which the criteria to have an
horizon are known: in the original coordinate system, the existence of a black hole can be expressed
as the existence of a Killing horizon for a time-like Killing vector, covering the singularity.
The last observation, then, is that in the general case no time-like Killing vector exists. 
\par
To see this consider for instance the $\overline{\mathbb{CP}}^{1}$-model:
this model has only one complex scalar field $Z$ living
on the coset space $Sl(2;\mathbb{R})/SO(2)$ and associated K\"ahler potential $e^{\mathcal{K}}=1-|Z|^{2}$,
so that we have the constraint $0\leq |Z|^{2}< 1$.
Choosing $\mathtt{C}_{\Lambda}=(-2,0)$, the potential can be readily be calculated to be
\begin{equation}
  \label{eq:PotCPn}
  \mathtt{V} \; =\; 2g^{2}\ \left[
                       1\ +\ 2\ e^{\mathcal{K}}
                    \right]\; ,
\end{equation}
which is manifestly positive. 
Imposing $\mathcal{I}^{0}=0$ in order to have $\mathbb{R}^{3}$ as the base space, and $\mathcal{I}_{1}=0$
in order to have a static solution ({\em i.e.\/} $\omega =0$), the equations of motion imply that 
a simple solution is
\begin{equation}
  \label{eq:PotCPn1}
  \mathcal{I}_{0} \; =\; \frac{g\tau}{\sqrt{2}} \hspace{.5cm}\mbox{and}\hspace{.5cm}
  \mathcal{I}^{1} \; =\; \sqrt{2}\ g\lambda \hspace{.5cm}\longrightarrow\hspace{.3cm}
  \frac{1}{2|X|^{2}} \ =\ g^{2}\left[ \tau^{2} \ -\ \lambda^{2}\right] \; , 
\end{equation}
where $\lambda$ is a real constant. 
If $\lambda =0$ the above solution leads to $DS_{4}$, whereas if $\lambda\neq 0$ we can introduce
a new coordinate $t$ through $\tau =\lambda\cosh\left( gt\right)$, such that the solution is given 
by
\begin{eqnarray}
  \label{eq:PotCPn2}
  ds^{2} & =& dt^{2}\ -\ \sinh^{2}\left( gt\right)\ d\vec{x}^{2}_{(3)} \; ,\\
  Z     & =& -i\ \cosh^{-1}\left( gt\right) \; .
\end{eqnarray}
At late times the metric is $DS_{4}$ but is singular when $t=0$; at that point in time also the scalar
becomes problematic as $|Z(t=0)|^{2}=1$, violating the bound, which in its turn implies that the 
contribution of the scalars to the energy-momentum tensor blows up.
Returning to the point we were going to make, it is paramount that in this case no time-like Killing vector
exists. Had we on the other hand taken $\mathcal{I}^{1}= \sqrt{2}gp\ r^{-1}$, in which case a time-like
Killing vector exists, the metric can be transformed to the static form
\begin{equation}
  \label{eq:PotCPn3}
  ds^{2} \ =\ \textstyle{p^{2} + R^{2}-g^{2}R^{4}\over R^{2}}\ dt^{2}
         \ -\ \textstyle{R^{4}\over ( R^{2}+p^{2})(p^{2} + R^{2}-g^{2}R^{4}/4)}\ dR^{2}
         \ -\ R^{2}dS^{2} \; .
\end{equation}
This metric has one Killing horizon, identified with the cosmological horizon, for $R>0$ and is therefore a 
naked singularity, with the singularity being located at $R=0$. 
In the static coordinates, the scalar field reads $Z=-ip\ {\textstyle (p^{2}+R^{2})^{-1/2}}$, which explicitly breaks
the bound $0\leq |Z|^{2}< 0$ at $R=0$, showing once again the link between the regularity of the metric
and that of the scalars.
\par
A manageable prescription for deciding when a solution describes a black hole is clearly desirable.
In this respect, we would like to mention the isolated horizon formalism (see {\em e.g.\/}\cite{Ashtekar:2004cn})
which attempt to give a local definition of horizons, without a reference to the existence of time-like Killing
vectors. This formalism was recently applied to sugras in ref.~\cite{Liko:2007mu} and similar work
for fake sugras is in progress.
%%%%%%%%%%%%%%%%%%%%%%%%%%%%%%%%%%%%%%%%%%%%%%%%%%%%%%%%%%%%%%%%%%%%%%%%%%%%%%%%%%%%%%%%%%%%%%%%%%%%%%%%%%%%%%%%%
\section{Null case}
\label{sec:Null}
%%%%%%%%%%%%%%%%
In this section we shall characterise the fake-supersymmetric solutions in the so-called null-case, by which
is meant the case when $V^{2}=0$: for simplicity we shall restrict ourselves to the theories
with no YM-type couplings, a full analysis along the lines of ref.~\cite{Hubscher:2008yz} being possible
but, seeing the results obtained in that reference, not very rewarding.
As in the time-like case, the difference with the supersymmetric case lies in the fact that the vector-bilinear
$L$ to be introduced below, is not a Killing vector; introducing then an adapted coordinate $v$ through 
$L^{a}\partial_{a}=\partial_{v}$, we see that the metric will be explicitly $v$-dependent, unlike the
supersymmetric case. The aim of this section, then, is to determine this $v$-dependence and give
2 minimal and simple, yet generic, solutions showing the changes brought about by the $\mathbb{R}$-gauging.
\par
In the Null case, the norm of the vector $V$ vanishes, whence $X=0$. This means that the 2 spinors $\epsilon_{I}$
are parallel, and following refs.~\cite{Meessen:2006tu,Hubscher:2008yz}, 
we shall put $\epsilon_{I}=\phi_{I}\epsilon$, for some functions $\phi_{I}$ and
the independent spinor $\epsilon$. The decomposition of $\epsilon^{I}$ follows from its definition as $\epsilon^{I}=(\epsilon_{I})^{*}$,
which then implies $\epsilon^{I}=\phi^{I}\epsilon^{*}$, where we have defined $\overline{\phi_{I}}=\phi^{I}$.
Furthermore, without loss of generality we can normalise the $\phi$'s such that $\phi_{I}\phi^{I}=1$. 
Once we take into account this normalisation, we can write down a completeness relation for the $I$-indices which is
\begin{equation}
  \label{eq:21}
  \Delta_{I}{}^{J} \; =\; \phi_{I}\phi^{J} \; +\; \varepsilon_{IK}\Phi^{K}\ \varepsilon^{JL}\Phi_{L} \; ,
\end{equation}
which is such that $\Delta_{I}{}^{J}\phi_{J}=\phi_{I}$ and $\Delta_{I}{}^{J} \varepsilon_{JK}\phi^{K}=\varepsilon_{IK}\phi^{K}$.
Moreover one can see that $\overline{\Delta_{I}{}^{J}}=\Delta_{J}{}^{I}$.
\par
Projecting, then, the fKSEs (\ref{eq:20},\ldots ,\ref{eq:20d}) onto the $\phi$'s we obtain
\begin{eqnarray}
  \label{eq:22}
  0 & =& \mathbb{D}_{a}\epsilon \; +\; \phi^{I}\nabla_{a}\phi_{I}\, \epsilon \; , \\
  \label{eq:22b}
  0 & =& \left( \mathcal{T}_{ab}^{+}\ +\ \textstyle{ig\over 4}\mathtt{C}_{\Lambda}\mathcal{L}^{\Lambda}\ \eta_{ab}\right)\ \gamma^{b}\epsilon^{*}
         \; -\; \varepsilon^{IJ}\phi_{I}\nabla_{a}\phi_{J}\ \epsilon \; ,\\
  \label{eq:22c}
  0 & =& i\slashed{\partial}Z^{i}\ \epsilon^{*} \; , \\
  \label{eq:22d}
  0 & =& \left[ \slashed{G}^{i+} \; +\: \mathtt{W}^{i}\right]\ \epsilon \; .
\end{eqnarray}
In order to advance we will introduce an auxiliary spinor $\eta$, normalised by 
$\overline{\epsilon}\eta = \textstyle{1\over \sqrt{2}} = -\overline{\eta}\epsilon$;
due to the introduction of $\eta$ we can introduce 4 null-vectors
\begin{equation}
  \label{eq:23}
  \begin{array}{lclclcl}
    L_{a} & =& i\overline{\epsilon}\gamma_{a}\epsilon^{*} &\hspace{.3cm},\hspace{.3cm}& N_{a} & =& i\overline{\eta}\gamma_{a}\eta^{*} \; ,\\
    M_{a} & =& i\overline{\eta}\gamma_{a}\epsilon^{*} & ,& \overline{M}_{a} & =& i\overline{\epsilon}\gamma_{a}\eta^{*} \; ,
  \end{array}
\end{equation}
where $L$ and $N$ are real vectors and by construction $M^{*}=\overline{M}$, whence the notation.
Observe that eq.~(\ref{eq:Bil2a}) implies that the vector $L$ is nothing but $V$, but we shall denote it by $L$(ightlike)
in order to avoid confusion with the foregoing section.
Given the above definitions it is a straightforward yet tedious calculation to show that they form an ordinary normalised null-tetrad, 
{\em i.e.\/} the only non-vanishing contractions are
\begin{equation}
  \label{eq:1}
  L^{a}\ N_{a} \; =\; 1 \; =\; -\ M^{a}\ \overline{M}_{a} \;\;\;\mbox{which implies}\;\;\;
  \eta_{ab} \ =\ 2\ L_{(a}N_{b)} \; -\; 2\ M_{(a}\overline{M}_{b)} \; .
\end{equation}
\par
Apart from the vectors one can also define imaginary-self-dual 2-forms,
analogous to the ones defined in eq.~(\ref{eq:Bil3}), by
\begin{equation}
  \label{eq:2}
  \begin{array}{lclclcl}
    \Phi^{1}_{ab} & \equiv& \overline{\epsilon}\gamma_{ab}\epsilon &\hspace{.2cm},\hspace{.2cm}&
    \Phi^{1} & =& \sqrt{2}\ L\wedge \overline{M} \; , \\
     \Phi^{2}_{ab} & \equiv& \overline{\epsilon}\gamma_{ab}\eta &\hspace{.2cm},\hspace{.2cm}&
    \Phi^{2} & =& \textstyle{1\over \sqrt{2}}\left[ L\wedge N \ +\ M\wedge \overline{M}\right] \; , \\
     \Phi^{3}_{ab} & \equiv& \overline{\eta}\gamma_{ab}\eta &\hspace{.2cm},\hspace{.2cm}&
    \Phi^{3} & =& -\sqrt{2}\ N\wedge M \; , \\
  \end{array}
\end{equation}
where the identification on the r.h.s.~ follows from the Fierz identities.
\par
The introduction of the above auxiliary spinor is not unique, and there still is some freedom left;
first of all we have the freedom to rotate $\epsilon$ and $\eta$ by $\epsilon\rightarrow e^{i\theta}\epsilon$
and $\eta\rightarrow e^{-i\theta}\eta$. This rotation does not affect $L$ nor $N$, but rotates $M\rightarrow e^{-2i\theta}M$
and $\overline{M}\rightarrow e^{2i\theta}\overline{M}$: we will use this freedom to get rid of a phase-factor when
introducing a coordinate expression for the tetrad.
The second freedom arises, because a shift $\eta\rightarrow \eta +\delta\ \epsilon$, with $\delta$ a complex function,
does not affect the normalisation condition. The effect of this shift on the vectors is
\begin{equation}
  \label{eq:3}
  L\ \rightarrow\ L 
  \; ,\; M\ \rightarrow\ M \ +\ \delta\ L
  \; ,\; N\ \rightarrow\ N \ +\ |\delta|^{2}\ L \ +\ \delta\ \overline{M} \ +\ \bar{\delta}\ M \; , 
\end{equation}
and this freedom can also be used to restrict the coordinate expressions of the tetrad. 
\par
Let us start introducing a coordinate system by introducing a coordinate $v$ through $L^{\flat}\equiv L^{a}\partial_{a}=\partial_{v}$,
and using eq.~(\ref{eq:22}) to derive
\begin{equation}
 \label{eq:5}
   \nabla_{a}L_{b} \; =\; g\mathtt{C}_{\Lambda}A^{\Lambda}_{a}\ L_{b} \; ,
\end{equation}
whence $L$ is a recurrent null vector: this is the defining property of a space with holonomy $\mathrm{Sim}(2)$ 
(see ref.~\cite{Gibbons:2007zu} for more information) and the combination $g\mathtt{C}_{\Lambda}A^{\Lambda}$ is called the 
recurrence 1-form.
%Joder, esa gente tiene todo el desarollo, pero bien... soy tan bueno como Gibbons y Pope, que no esta mal.
Anti-symmetrising this expression we see that $dL= g\mathtt{C}_{\Lambda}A^{\Lambda}\wedge L$, which implies not only $\mathtt{C}_{\Lambda}F^{\Lambda}\wedge L =0$,
but also $L\wedge dL =0$. This last result states that the vector $L$ is hyper-surface orthogonal, which implies the local existence of functions
$Y$ and $u$ such that $L = Ydu$.
Seeing, however, that $L$ is charged under the $\mathbb{R}$-symmetry, we can always gauge-transform the function $Y$ away, leaving
the statement that $L=du$, whence also that $\mathtt{C}_{\Lambda}A^{\Lambda} = \Upsilon\ L$, for some function $\Upsilon$. We can then write eq.~(\ref{eq:5})
as 
\begin{equation}
  \label{eq:6}
  \nabla_{a}\ L_{b} \; =\; g\Upsilon\ L_{a}L_{b} \;\;\;\mbox{which immediately implies}\;\; \nabla_{L}L \ =\ 0 \; ,
\end{equation}
so that $L$ is a geodesic null-vector. 
Given this information and the normalisation of the tetrad we can choose coordinates $u$, $v$, $z$ and $\bar{z}$
such that\footnote{
  See appendix (\ref{sec:NullCurv}) for the spin-connection and curvatures for this tetrad.
}
\begin{equation}
  \label{eq:8}
  \begin{array}{lclclcl}
    L & =& du   &\;\; ,\;\; & L^{\flat} & =& \partial_{v} \; ,   \\
    N & =& dv \ +\ Hdu \ +\varpi dz\ + \overline{\varpi}d\bar{z} & ,& N^{\flat} & =& \partial_{u} \ -\ H\partial_{v} \; ,\\
    M & =& e^{U}dz  & ,& M^{\flat} & =& -e^{-U}\left( \partial_{\bar{z}} \ -\ \overline{\varpi}\partial_{v}\right) \; ,\\
    \overline{M} & =& e^{U}d\bar{z} & ,&  \overline{M}^{\flat} & =& -e^{-U}\left( \partial_{z} \ -\ \varpi\partial_{v}\right) \; ,
  \end{array}
\end{equation}
where we used the $U(1)$-rotation $M\rightarrow e^{-2i\theta}M$ to get rid of a possible phase in the expression of $M$ and $\overline{M}$.
The spin-connection and curvatures for the tetrad is given in Appendix (\ref{sec:NullCurv}).
A last implication of the Fierz identities is that 
\begin{equation}
  \label{eq:4}
  \varepsilon_{(4)} \;\equiv\; \textstyle{1\over 4!}\ \varepsilon_{abcd}\ e^{a}\wedge e^{b}\wedge e^{c}\wedge e^{d}
                   \; =\; i\ L\wedge N\wedge M\wedge \overline{M} 
                   \; =\; i\ e^{+}\wedge e^{-}\wedge e^{\bullet}\wedge e^{\bar{\bullet}} \; ,
\end{equation}
which implies that $\varepsilon^{+-\bullet\bar{\bullet}}=i$.
\par
Given the above expressions for the tetrad one can calculate the implications of the restriction (\ref{eq:6}); one finds
\begin{equation}
  \label{eq:9}
  \partial_{v}H \; =\; g\Upsilon  \hspace{.3cm}\mbox{and}\hspace{.3cm} 0\ =\ \partial_{v}U \ =\ \partial_{v}\varpi\ =\ \partial_{v}\overline{\varpi} \; ,
\end{equation}
whence the only $v$-dependence of the metric resides in $H$;
The resulting form of the metric is called a Walker metric, in honour of the late A.G.~Walker, who was the first to give the general $d$-dimensional
metric of a space with holonomy contained in $\mathrm{Sim}(d-2)$ in ref.~\cite{art:walker1950}. 
\par
In order to determine $\Upsilon$, we can use the identity 
$\mathtt{C}_{\Lambda}F^{\Lambda} = d\left( \mathtt{C}_{\Lambda}A^{\Lambda}\right) = d\Upsilon\wedge L$, which presupposes knowing $F^{\Lambda}$.
\par
The generic form of $F^{\Lambda}$ can be derived from the fKSEs (\ref{eq:22b},\ref{eq:22d}): 
consider first of all eq.~(\ref{eq:22b}). 
Contraction with $i\overline{\epsilon}$ and $i\overline{\eta}$ leads to
\begin{eqnarray}
  \label{eq:10}
  \imath_{L}\mathcal{T}^{+} & =& \textstyle{ig\over 4}\ \mathtt{C}_{\Lambda}\mathcal{L}^{\Lambda}\ L \; ,\\
  \label{eq:10a}
  \imath_{M}\mathcal{T}^{+} & =& \textstyle{ig\over 4}\ \mathtt{C}_{\Lambda}\mathcal{L}^{\Lambda}\ M 
       \, +\, \textstyle{i\over \sqrt{2}} \phi_{I}\varepsilon^{IJ}\ d\phi_{J} \; .
\end{eqnarray}
Coupling the above information to the fact that as $\mathcal{T}^{+}$ is an imaginary-self-dual 2-form it must 
be expressible in terms
of the $\Phi$'s defined in eq.~(\ref{eq:2}), we see that 
\begin{equation}
  \label{eq:11}
  \mathcal{T}^{+} \; =\; \aleph\ L\wedge\overline{M} \ -\ 
    \textstyle{ig\over 4}\ \mathtt{C}_{\Lambda}\mathcal{L}^{\Lambda}\ \left[ L\wedge N + M\wedge \overline{M}\right] 
  \;\;\mbox{with}\;\;
  \sqrt{2}\aleph \ =\  i\phi_{I}\varepsilon^{IJ}\nabla_{N}\phi_{J} \; , 
\end{equation}
and furthermore
\begin{equation}
  \label{eq:13}
  \sqrt{2}\phi_{I}\varepsilon^{IJ}\nabla_{\overline{M}}\phi_{J} \; =\; g\mathtt{C}_{\Lambda}\mathcal{L}^{\Lambda}
  \;\;\; ,\;\;\; 0 \; =\; \phi_{I}\varepsilon^{IJ}\nabla_{L}\phi_{J} \; =\; \phi_{I}\varepsilon^{IJ}\nabla_{M}\phi_{J} \; .
\end{equation}
Giving eq.~(\ref{eq:22d}) a similar treatment leads to 
\begin{equation}
  \label{eq:14}
  G^{i+} \; =\; \aleph^{i}\ L\wedge\overline{M} \, -\, \textstyle{1\over 4}\ \mathtt{W}^{i}
   \left[ L\wedge N \, +\, M\wedge\overline{M}\right] \; ,
\end{equation}
where $\aleph^{i}$ are, at this point, undetermined functions. Using the by-now-well-known rule 
$F^{\Lambda +}=i\overline{\mathcal{L}}^{\Lambda}\mathcal{T}^{+}+2f_{i}^{\Lambda}G^{i+}$, we find that 
\begin{equation}
  \label{eq:15}
  F^{\Lambda +} \ =\ \varphi^{\Lambda}\ L\wedge\overline{M} 
    \, +\, V^{\Lambda}\ \left[ L\wedge N + M\wedge\overline{M}\right] \; ,
\end{equation}
where we introduced
\begin{equation}
  \label{eq:DefVLambda}
  V^{\Lambda} = \textstyle{g\over 8}
             \left(\ 
                4\overline{\mathcal{L}}^{\Lambda}\mathcal{L}^{\Sigma} + \mathrm{Im}(\mathcal{N})^{-1|\Lambda\Sigma}\ 
             \right)\mathtt{C}_{\Sigma}
\end{equation}
and 
\begin{equation}
  \label{eq:17}
  \aleph  = 2i\ \mathcal{L}_{\Lambda}\ \varphi^{\Lambda}\; ;\; 
  \aleph^{i} \; =\; -\bar{f}_{\Lambda}^{i}\ \varphi^{\Lambda} \;\; \longleftrightarrow\;\; 
  \varphi^{\Lambda} \; =\; i\aleph\ \overline{\mathcal{L}}^{\Lambda} 
        \ +\  2\aleph^{i}\ f_{i}^{\Lambda} \; .
\end{equation}
Using then $F^{\Lambda} = F^{\Lambda +}+F^{\Lambda -}= 2\mathrm{Re}\left( F^{\Lambda +}\right)$, and doing
the comparison $d\Upsilon\wedge L=\mathtt{C}_{\Lambda}F^{\Lambda}$, we obtain
\begin{eqnarray}
  \label{eq:18}
  \nabla_{L}\Upsilon & =& -\mathtt{C}_{\Lambda}\left[ V \ +\ \overline{V}\right]^{\Lambda} \; ,\\
  \label{eq:18b}
  \nabla_{M}\Upsilon & =& \mathtt{C}_{\Lambda}\ \varphi^{\Lambda} \; ,\\
  \label{eq:18c}
  \nabla_{\overline{M}}\Upsilon & =& \mathtt{C}_{\Lambda}\ \overline{\varphi}^{\Lambda} \; .
\end{eqnarray}
It is clear that eq.~(\ref{eq:18}) is the key to the possible $v$-dependence: 
in order to integrate it and obtain $H$ through eq.~(\ref{eq:9}),
we need to know the coordinate dependence of the scalars $Z$.
\par
Information about said coordinate dependence can of course be obtained from eq.~(\ref{eq:22c}), by contracting it with the $i\overline{\epsilon}$
and $i\overline{\eta}$. The result is that
\begin{equation}
  \label{eq:19}
  0\, =\, \nabla_{L} Z^{i} \, =\, \partial_{v}Z^{i} \;\;\;\mbox{and}\;\;\; 
  0\, =\, \nabla_{M} Z^{i} \, =\, e^{-U}\ \partial_{\bar{z}}Z^{i} \; ,
\end{equation}
so that the $Z^{i}$ depend only on $u$ and $z$. Likewise, the $\overline{Z}^{\bar{\imath}}$ depend only on $u$ and $\bar{z}$.
% In the supersymmetric case, where $L$ is a Killing vector, it is desireable to have scalars that are $v$-independent as to avoid
% complicated transformations for the scalars in order to compensate for translations in $v$; in that case said $v$-independence 
% is achieved by gauge-fixing $\imath_{L}A^{\Lambda}=0$. Now, even though it may not be obvious from the way we are writing the 
% splitting of the gauge fields in sectors, the same gauge-fixing is attainable, in fact the gauge field of the $\mathbb{R}$-symmetry,
% {\em i.e.\/} $\mathtt{C}_{\Lambda}A^{\Lambda}$, is already in said gauge.
% The implication of this gauge-fixing is, then, the same as in the supersymmetric case: the scalars can be taken to be $v$-independent.
% Observe that the need to gauge-fix is only needed for the non-Abelian sectors as eq.~(\ref{eq:19}) for scalars in the Abelian sector
% immediately implies that they are $v$- and $\bar{z}$-independent.
\par
Using the fact that the scalars are $v$-independent, integration of eq.~(\ref{eq:18}) is straightforward and leads to
\begin{eqnarray}
  \label{eq:7}
  \Upsilon & =& -\textstyle{g\over 4}\left[
                      4\left|\mathtt{C}_{\Lambda}\mathcal{L}^{\Lambda}\right|^{2}
                      \ +\ \mathrm{Im}\left(\mathcal{N}\right)^{-1|\Lambda\Sigma}\mathtt{C}_{\Lambda}\mathtt{C}_{\Sigma} 
                  \right]\ v \, +\, \Upsilon_{1}(u,z,\bar{z}) \; , \\
  \label{eq:7a}
  H & =& -\textstyle{g^{2}\over 8}\left[
                      4\left|\mathtt{C}_{\Lambda}\mathcal{L}^{\Lambda}\right|^{2}
                      \ +\ \mathrm{Im}\left(\mathcal{N}\right)^{-1|\Lambda\Sigma}\mathtt{C}_{\Lambda}\mathtt{C}_{\Sigma} 
                  \right]\ v^{2} \, +\, \Upsilon_{1}\ v \ +\ \Upsilon_{0}(u,z,\bar{z}) \; .
\end{eqnarray}
By doing a coordinate transformation $v\rightarrow v + f(u,z,\bar{z})$ we can take $\Upsilon_{1}=0$, but for the moment
we shall ignore this possibility. 
\par
$H$ can be written in terms of the potential $\mathtt{V}$ in eq.~(\ref{eq:Potential}), with $\mathtt{P}_{\Lambda}=0$
as we are ignoring possible non-Abelian couplings, as
\begin{equation}
  \label{eq:7b}
  H \ =\ \textstyle{1\over 2}\left[ 
         g^{2}\left|\mathtt{C}_{\Lambda}\mathcal{L}^{\Lambda}\right|^{2} \ -\ \mathtt{V}
         \right]\ v^{2} \ +\ \Upsilon_{1}v \ +\ \Upsilon_{0} \; ,
\end{equation}
which is calculationally advantageous when $\mathtt{V}$ is known.
\par
At this point we have nearly completely specified the $v$-dependence of the solution, the only field missing
being the $A^{\Lambda}$; in order to determine its $v$-dependence it is worthwhile to impose the gauge-fixing
$\imath_{L}A^{\Lambda}=0$, which is always possible and is furthermore consistent with the earlier result 
$\mathtt{C}_{\Lambda}A^{\Lambda}=\Upsilon\ L$.
As a result of this gauge fixing we have that 
\begin{equation}
  \label{eq:7c}
  \partial_{v}A^{\Lambda} \ =\ \pounds_{L}A^{\Lambda} 
       \ =\  d\left(\imath_{L}F^{\Lambda}\right)
       \ =\  -\left(\ V\ +\ \overline{V}\ \right)^{\Lambda}\, L \; ,
\end{equation}
so that 
\begin{equation}
  \label{eq:7d}
  A^{\Lambda} \; =\; -\left(\ V\ +\ \overline{V}\ \right)^{\Lambda}\, v\ L \; +\; \tilde{A}^{\Lambda} 
                   \; =\; \textstyle{g\over 4}\ \mathsf{F}^{-1|\Lambda\Sigma}\mathtt{C}_{\Sigma}\ v\ L \; +\; \tilde{A}^{\Lambda} \; ,
\end{equation}
where $\tilde{A}^{\Lambda}$ is a $v$-independent 1-form satisfying $\imath_{L}\tilde{A}^{\Lambda}=0$, and
$\mathsf{F}$ is the imaginary part of the prepotential's Hessian; see eq.~(\ref{eq:PrePotImNinv}) for why this ocurrs..
Given this expression for the vector potentials, the Bianchi identity is automatically satisfied, 
but, as in the time-like case, this does not necessarily mean that any $\tilde{A}^{\Lambda}$
leads to a field-strength of the desired form. Calculating the comparison we find that
\begin{eqnarray}
  \label{eq:7e}
  d\tilde{A}^{\Lambda} & =& \left( V-\overline{V}\right)^{\Lambda}\ M\wedge\overline{M} \nonumber \\
      & +& \left( \phi^{\Lambda}+\theta_{M}\left[ v\left( V+\overline{V}\right)^{\Lambda}\right]\right)
             \ L\wedge\overline{M}
      \ +\ \left( \overline{\phi}^{\Lambda}+\theta_{\overline{M}}\left[ v\left( V+\overline{V}\right)^{\Lambda}\right]\right)
             \ L\wedge M \, .
\end{eqnarray}
\par
Let us at this point return to the fKSEs, and evaluate eq.~(\ref{eq:20c}) using eqs.~(\ref{eq:14}) and (\ref{eq:19}). This
evaluation results in 
\begin{equation}
  \label{eq:26}
  i\theta_{+}Z^{i}\ \gamma^{+}\epsilon^{I} \ +\ i\theta_{\bullet}Z^{i}\ \gamma^{\bullet}\epsilon^{I} \; =\;
    -\varepsilon^{IJ}\left[ \mathtt{W}^{i}\gamma^{-}\ -\ 2\alpha^{i}\gamma^{\bar{\bullet}}\right]\ \gamma^{+}\epsilon_{J} \; .
\end{equation}
The above equation is readily seen to be solved by observing that the constraint $\gamma^{+}\epsilon_{J}=0$
not only leads to $\gamma^{+}\epsilon^{I}=0$ under complex conjugation, but also to $\gamma^{\bar{\bullet}}\epsilon_{I}=0$ and
$\gamma^{\bullet}\epsilon^{I}=0$; 
these last implications are due to the fact that we dealing with chiral spinors and the normalisation in eq.~(\ref{eq:4}).
\par
Doing a similar analysis on the fKSE (\ref{eq:20}) in the $v$-direction shows that the spinor $\epsilon_{I}$, whence also $\epsilon^{I}$,
is $v$-independent. The other equations become
\begin{eqnarray}
  \label{eq:27}
  \mathbb{D}_{\bar{\bullet}}\epsilon_{I} & =& 0\; ,\\
  \label{eq:27a}
  \mathbb{D}_{\bullet}\epsilon_{I} & =& \textstyle{ig\over 2}\ \mathtt{C}_{\Lambda}\mathcal{\Lambda}\ \varepsilon_{IJ}\gamma^{\bar{\bullet}}\epsilon^{J} \; , \\
  \label{eq:27c}
  \mathbb{D}_{+}\epsilon_{I} & =& -\aleph\ \varepsilon_{IJ}\gamma^{\bar{\bullet}}\epsilon^{J} \; .
\end{eqnarray}
\par
Using the definition (\ref{eq:16}) and the spin-connection in eq.~(\ref{eq:NCspincon}), we can expand eqs.~(\ref{eq:27}) and (\ref{eq:27a}) as
\begin{eqnarray}
  \label{eq:28}
  0 & =& \theta_{\bar{\bullet}}\epsilon_{I} \ -\ \textstyle{1\over 2}\theta_{\bar{\bullet}}\left( U +\textstyle{1\over 2}\mathcal{K}\right)\ \epsilon_{I} \; , \\
  \label{eq:28a}
  0 & =& \theta_{\bullet}\epsilon_{I} \ +\ \textstyle{1\over 2}\theta_{\bullet}\left( U +\textstyle{1\over 2}\mathcal{K}\right)\ \epsilon_{I} 
          \ -\ \textstyle{ig\over 2}\mathtt{C}_{\Lambda}\mathcal{L}^{\Lambda}\ \varepsilon_{IJ}\gamma^{\bar{\bullet}}\epsilon^{J} \; ,
\end{eqnarray}
The first equation is easily integrated by putting
\begin{equation}
  \label{eq:29}
  \epsilon_{I} \; =\; \exp\left( \textstyle{1\over 2}\ S\right)\ \chi_{I} (u,z) \hspace{.5cm}\mbox{with}\hspace{.5cm}
  S \ \equiv\ U \ +\ \textstyle{1\over 2}\mathcal{K} \; ,
\end{equation}
which upon substitution into eq.~(\ref{eq:28a}) leads to
\begin{equation}
  \label{eq:30}
  \partial_{z}\chi_{I} \ +\ \left(\partial_{z}S\right)\chi_{I} \; =\; \textstyle{ig\over 2}\ \mathtt{C}_{\Lambda}\mathcal{X}^{\Lambda}\
       \varepsilon_{IJ}\gamma^{\bar{\bullet}}\ e^{S}\chi^{J} \; .
\end{equation}
This last equation is potentially dangerous as it has a residual $\bar{z}$-dependence, even though $\eta$ and $\mathcal{X}^{\Lambda}$
are $\bar{z}$-independent; it is this possible inconsistency that fixes $S$, 
as can be seen by deriving eq.~(\ref{eq:30}) w.r.t.~$\bar{z}$ and using the complex conjugated version of eq.~(\ref{eq:30})
to get rid of $\eta^{I}$ in the resulting equations. The result is that $S$ has to satisfy
\begin{equation}
  \label{eq:31}
  \partial_{z}\partial_{\bar{z}}S \; =\; -\textstyle{g^{2}\over 2}\ e^{2S}\ \left|\mathtt{C}_{\Lambda}\mathcal{X}^{\Lambda}\right|^{2} 
  \;\;\;\longrightarrow\;\;\;
  e^{-2S} \, =\, \textstyle{g^{2}\over 2}\ \left|\mathtt{C}_{\Lambda}\mathcal{X}^{\Lambda}\right|^{2}\ \left( 1+|z|^{2}\right)^{2} \; .
\end{equation}
This unique choice for $S$ is a necessary condition for the eqs.~(\ref{eq:28}) and (\ref{eq:28a}) to admit a solution, but it may
not be sufficient; in the next section we shall discuss the simplest null-case solution to the minimal theory, and show that the 
system can be solved completely. The lesson to be learned from that section is that the system (\ref{eq:28},\ref{eq:28a}) once
we introduce $S$, corresponds to an equation determining spinors on a 2-sphere, 
and has solutions even though this is hard to see.
% \par
% The $\mathtt{C}_{\Lambda}\mathcal{X}^{\Lambda}$- factor can be absorbed by redefining
% $\chi_{I}\ =\ \sqrt{\mathtt{C}_{\Lambda}\mathcal{X}^{\Lambda}}\ \eta_{I}$ which converts eq.~(\ref{eq:31}) into
% \begin{equation}
%   \label{eq:31a}
%   \partial_{z}\eta_{I} \ +\ \left(\partial_{z}\tilde{S}\right)\eta_{I} \, =\, \textstyle{ig\over 2}\ 
%        \varepsilon_{IJ}\gamma^{\bar{\bullet}}\ e^{\tilde{S}}\eta^{J} 
%   \hspace{.6cm}\mbox{with}\hspace{.4cm}
%   e^{-\tilde{S}} \ =\ \textstyle{g\over\sqrt{2}}\left( 1+|z|^{2}\right) \; , 
% \end{equation}
% which is just the spinor equation on $S^{2}$ in stereographic coordinates.
%%%%%%%%%%%%%%%%%%%%%%%%%%%%%%%%%%%%%%%%%%%%%%%%%%%%%%%%%%%%%%%%%%%%%%%%%%%%%%%%%
\subsection{The electrically charged Nariai cosmos belongs to the Null case}
\label{sec:NullSols}
%%%%%%%%%%%%%%%%%%%%%%%%%%%%%%%%%%%%%%%%%%%%%%%%%%%%%%%%%%%%%%%%%%%%%%%%%%%%%%%%%
The minimal theory is obtained by putting $\mathcal{V}^{T}=(1,-i/2)$, which leads to
the monodromy matrix $\mathcal{N}=-i/2$, so that $\mathrm{Re}(\mathcal{N})=0$.
if we then further fix $\mathtt{C}_{0}=2$, we see that the minimal De Sitter theory
is given by
\begin{equation}
  \label{eq:MinDSaction}
  \int_{4}\sqrt{g}\left( R\ -\ F^{2}\ -\ 6g^{2}\right) \; .
\end{equation}
Using the general results obtained thus far, we can write down the following solution
\begin{eqnarray}
  \label{eq:NullNariai}
  ds^{2} & =& 2du\left( dv -g^{2}v^{2}du\right) \, -\, \frac{dzd\bar{z}}{g^{2}(1+|z|^{2})^{2}} \; .\nonumber \\
  A     & =& -gv\ du\; ,
\end{eqnarray}
A small analysis shows that the metric is nothing more than $DS_{2}\times S^{2}$, albeit in a non-standard coordinate system, and the solutions is known to the literature as the 
electrically charged Nariai solution \cite{art:nariai}.
Observe that the local holonomy of the Nariai solution is not the full $\mathfrak{sim}(2)$,
but rather 
$\mathfrak{so}(1,1)\oplus\mathfrak{so}(2)\subset \mathfrak{sim}(2)$ \cite{Gibbons:2007zu}.
\par
In order to discuss the preserved fake-supersymmetries it is easier to write the metric as
\begin{equation}
  \label{eq:NullNariai1}
  ds^{2} \; =\; 2du\left( dv -g^{2}v^{2}du\right) 
        \, -\, \frac{1}{4g^{2}}\left[ d\theta^{2}\ +\ \sin^{2}(\theta )d\varphi^{2}\right] \; ,
\end{equation}
and consider the fake-supergravity equations in terms of a 2-component vector of Majorana spinors,
also denoted by $\epsilon$, namely
\begin{equation}
  \label{eq:NullNariai2}
  \nabla_{a}\epsilon \ -\ gA_{a}\epsilon 
       \; =\; -\textstyle{1\over 4}\slashed{F}\gamma_{a}\sigma^{2}\epsilon
              -\textstyle{g\over 2}\gamma_{a}\sigma^{2}\epsilon \; .
\end{equation}
The solution to the above equation is then seen to be 
\begin{equation}
  \label{eq:NullNariai3}
  \epsilon \; =\; \exp\left( \textstyle{\theta\over 2}\gamma^{3}\sigma^{2}\right)\
                  \exp\left( -\textstyle{\varphi\over 2}\gamma^{34}\right)\ \epsilon_{0}
  \hspace{.4cm}\mbox{with}\hspace{.4cm} \gamma^{+}\epsilon_{0}=0\; ,
\end{equation}
where $\epsilon_{0}$ is a 2-vector of constant spinors. 
Some remarks are in order: in supersymmetry one can associate a Lie superalgebra to a given supersymmetric
solution \cite{Gauntlett:1998kc},
and for the supersymmetric $aDS_{2}\times S^{2}$ maximally supersymmetric solutions in minimal 
$N=2$ $d=4$, this algebra is $\mathfrak{su}(1,1|2)$. In the fake-supersymmetric case, however, one cannot
assign a Lie superalgebra to the solution, as the vector bilinears which would represent the supertranslation
part, do not lead to Killing vectors; this fact is already illustrated by eq.~(\ref{eq:5}).
A perhaps worrisome point is the action of the De Sitter's Killing vectors on the preserved fake-supersymmetry,
especially since the Killing spinors are $u$- and $v$-independent. Taking into account that the
spinors are gauge-dependent objects means that this action is defined using the $\mathbb{R}$-covariant Lie 
derivative on spinors \cite{Ortin:2002qb}; 
this derivative is defined for Killing vectors $X$ and $Y$ as
\begin{equation}
  \label{eq:Kosmann}
  \mathbb{L}_{X}\epsilon \; =\; \nabla_{X}\epsilon 
         \ +\ \textstyle{1\over 4}\left(\partial_{a}X_{b}\right)\ \gamma^{ab}\epsilon
         \ -\ g\xi_{X}\ \epsilon \;\;\;\mbox{with}\;\;\;
  \left\{
    \begin{array}{rcl}
       d\xi_{X} & =& \pounds_{X}A \\
       \xi_{[X,Y]} & =& \pounds_{X}\xi_{Y}-\pounds_{Y}\xi_{X}
    \end{array}
  \right.
\end{equation}
Using this Lie derivative, one can see that $\mathbb{L}_{X}\epsilon =0$ for any $X\in\mathrm{Isom}(DS_{2})$.
%%%%%%%%%%%%%%%%%%%%%%%%%%%%%%%%%%%%%%%%%%%%%%%%%%%%%%%%%%%%%%%%%%%%%%%%%%%%%%%%%%%%%%%%%%%%%%%%%%%%%%%%%%%%%%%%%
\subsection{Holomorphic scalars and deformations of the Nariai cosmos}
\label{sec:Holomorphic}
%%%%%%%
In the supersymmetric case, there are 2 generic classes of solutions in the null case whose supersymmetry
is straightforward to see: the first are the pp-waves which are characterised by the fact that the scalars
depend only on $u$, and the {\em cosmic strings} which are characterised by vanishing vector potentials $A^{\Lambda}$,
vanishing Sagnac connection, $\varpi =0$,
and a holomorphic spacetime dependence of the scalars, {\em i.e.\/} $Z^{i}=Z^{i}(z)$ \cite{Tod:1995jf,Meessen:2006tu}.
In this section we will consider the analogue of the latter case and impose $\varpi =0$ and 
that $Z^{i}$ is a function of $z$ only.
Due to eq.~(\ref{eq:7d}), however, the vector potentials cannot vanish and we will look
for the minimal expression for $\tilde{A}^{\Lambda}$ for which the Bianchi identity, eq.~(\ref{eq:7e})
is solved: minimality implies that $\phi^{\Lambda} = v e^{-U}\partial_{\bar{z}}\left( V+\overline{V}\right)^{\Lambda}$
and the Bianchi identity reduces to
\begin{equation}
  \label{eq:Hol1}
  d\tilde{A}^{\Lambda} \; =\; 2i\ \mathrm{Im}\left(\frac{\mathcal{X}^{\Lambda}}{g\mathtt{C}_{\Sigma}\mathcal{X}^{\Sigma}}\right)
     \, \frac{dz\wedge d\bar{z}}{(1+|z|^{2})^{2}} \; ,
\end{equation}
a solution to which exists locally and determines $\tilde{A}_{u}=0$ and $\tilde{A}_{z}^{\Lambda}$ and 
$\tilde{A}^{\Lambda}_{\bar{z}}$ as functions of $z$ and $\bar{z}$.
\par
Given the above identifications we can use eq.~(\ref{eq:15}) to calculate the constraints imposed by the 
Maxwell e.o.m.s, {\em i.e.\/} $\mathcal{B}_{\Lambda}=0$ in eq.~(\ref{eq:VectEOM2}), which leads to
\begin{eqnarray}
  \label{eq:Hol2a}
  \mathcal{N}_{\Lambda\Sigma}\ \partial_{z}\left( V +\overline{V}\right)^{\Sigma} & =&
  \partial_{z}\left[ \overline{\mathcal{N}}_{\Lambda\Sigma}V^{\Sigma}
                \ +\ \mathcal{N}_{\Lambda\Sigma}\overline{V}^{\Sigma}
              \right] \; ,\\
  \label{eq:Hol2b}
  \overline{\mathcal{N}}_{\Lambda\Sigma}\ \partial_{\bar{z}}\left( V +\overline{V}\right)^{\Sigma} & =&
  \partial_{\bar{z}}\left[ \overline{\mathcal{N}}_{\Lambda\Sigma}V^{\Sigma}
                \ +\ \mathcal{N}_{\Lambda\Sigma}\overline{V}^{\Sigma}
              \right] \; , \\
  \label{eq:Hol2c}
  \partial_{z}\left[
                \overline{\mathcal{N}}_{\Lambda\Sigma}\ \partial_{\bar{z}}\left( V +\overline{V}\right)^{\Sigma}
              \right]
      & =& 
  \partial_{z}\left[
                \mathcal{N}_{\Lambda\Sigma}\ \partial_{z}\left( V +\overline{V}\right)^{\Sigma}
              \right] \; ,
\end{eqnarray}
the contribution due to $\tilde{A}^{\Lambda}$ dropping out identically.
As eq.~(\ref{eq:Hol2b}) is the complex conjugated version of (\ref{eq:Hol2a}), and eq.~(\ref{eq:Hol2c})
is the integrability condition for eqs.~(\ref{eq:Hol2a}) and (\ref{eq:Hol2b}), we only need to 
see that eq.~(\ref{eq:Hol2a}) holds.
\par
Using the holomorphicity of the scalars in order to write $\partial_{z}=\partial_{z}Z^{i}\ \partial_{i}$,
one can rewrite eq.~(\ref{eq:Hol2a}) as an equation in Special Geometry, namely
\begin{eqnarray}
  \label{eq:Hol3a}
  \partial_{i}\overline{\mathcal{N}}_{\Lambda\Sigma}\ V^{\Sigma} \ +\
  \partial_{i}\mathcal{N}_{\Lambda\Sigma}\ \overline{V}^{\Sigma}
    & =& 2i\mathrm{Im}\left(\mathcal{N}\right)_{\Lambda\Sigma}\ \partial_{i}V^{\Sigma} \nonumber \\
    & =& gi\overline{\mathcal{L}}_{\Lambda}\ \mathtt{C}_{\Gamma}f_{i}^{\Gamma}
         \ -\ \textstyle{gi\over 4}\partial_{i}\mathrm{Im}\left(\mathcal{N}\right)_{\Lambda\Sigma}\,
              \mathrm{Im}\left(\mathcal{N}\right)^{-1|\Sigma\Gamma}\mathtt{C}_{\Gamma} \; .
\end{eqnarray}
Some straightforward algebra using the expressions (\ref{eq:SGdifiN}) and (\ref{eq:SGdifbiN}) shows that
the above equation holds, whence the Maxwell equations are solved for arbitrary scalar functions $Z^{i}(z)$.
\par
Had we been sure of the fact that the generic expressions for the fields we are using solve the 
fKSEs, we would have deduced from the KSIs that we only need to verify $\mathcal{B}_{++}=0$ as
to be sure that the proposed configuration solves the equations of motion. As we are not 100{\%} sure of this
fact, however, we checked that all of the equations of motion are indeed satisfied. 
As was to be expected from the discussion of the Maxwell equations, all the e.o.m.s reduce to 
Special Geometry calculations.
\par
In conclusion then, given an expression for $Z^{i}=Z^{i}(z)$, we need to find the local expression for
$\tilde{A}^{\Lambda}$ from eq.~(\ref{eq:Hol1}), and the solution is given by
\begin{eqnarray}
  \label{eq:NarGen1}
  ds^{2} & =& 2du\left( dv - \textstyle{1\over 2}H_{0}v^{2}\ du\right) \ -\ 
              \frac{4}{g^{2}\left|\mathtt{C}_{\Lambda}\mathcal{L}^{\Lambda}\right|^{2}}
              \frac{dzd\bar{z}}{(1+|z|^{2})^{2}} \; , \\
  & & \nonumber \\
  \label{eq:NarGen2}
  A^{\Lambda} & =& \textstyle{g\over 4}\mathsf{F}^{-1|\Lambda\Sigma}\mathtt{C}_{\Sigma}\ v\ du
                     \ +\ \tilde{A}^{\Lambda} \; ,
\end{eqnarray}
where
\begin{equation}
  \label{eq:NarGen3}
  H_{0} \; =\; \mathtt{V} \ -\ g^{2}\left| \mathtt{C}_{\Lambda}\mathcal{L}^{\Lambda}\right|^{2} \; .
\end{equation}
\par
Nariai-like solutions can be obtained by taking the scalars $Z^{i}$ to be constants, in which case
the $z\bar{z}$-part of the metric describes a 2-sphere of radius $g|\mathtt{C}_{\Lambda}\mathcal{L}|$.
Depending on $H_{0}$, the $uv$-part of the metric describes $DS_{2}$ ($H_{0}>0$), 2-dimensional Minkowski 
space ($H_{0}=0$) or $aDS_{2}$ ($H_{0}<0$).
As before, these spaces have local holonomy contained in $\mathrm{sim}(2)$; the solution for generic $Z^{i}(z)$, however,
has proper $\mathrm{sim}(2)$ holonomy. 
%%%%%%%%%%%%%%%%%%%%%%%%%%%%%%%%%%%%%%%%%%%%%%%%%%%%%%%%%%%%%%%%%%%%%%%%%%%%%%%%%%%%%%%%%%%%%%%%%%%%%%%%%%%%%%%%%
\section{Non-BPS solutions to $N=2$ sugra from fEYM}
\label{sec:PotIsNul}
%%%%%%%
As is well-known, there are models in $N=2$ $d=4$ sugra coupled to vector-multiplets for which one can choose the 
Fayet-Iliopoulos terms such that the hyper-multiplet contribution to the potential vanishes
(see {\em e.g.\/} \cite{Cremmer:1984hj} or \cite[sec.~9]{Andrianopoli:1996cm} for a discussion of this point).
As we are basically dealing with a Wick-rotated version of the general supersymmetric set-up, this implies that there
are fake-supersymmetric models in which the only contribution to the potential comes from the gauging of
the isometries, as the FI-contributions cancel. In that case the bosonic action (\ref{eq:VectAct}) coincides with that
of an ordinary YMH-type of supergravity theory, and we must conclude that for those specific models the solutions
we obtained are in fact non-BPS solutions of a regular supergravity theory.\footnote{
  Needless to say, this reasoning also holds for the ordinary gauged $N=2$ $d=4$ supergravities with potentials
  whose FI-contribution vanishes. 
} 
Let us illustrate this fact with an example: the dimensional reduction of minimal 5-dimensional sugra.
\par
The dimensional reduction of minimal 5-dimensional sugra leads to a specific $N=2$ $d=4$ sugra,
namely minimal sugra coupled to one vector-multiplet with a prepotential given by
\begin{equation}
  \label{eq:24}
  \mathcal{F}\left( \mathcal{X}\right) \; =\; -\textstyle{1\over 8}\ \frac{\left(\mathcal{X}^{1}\right)^{3}}{\mathcal{X}^{0}} \;  .
\end{equation}
With the usual choice $Z=\mathcal{X}^{1}/\mathcal{X}^{0}$, one finds that the scalar-manifold is $\mathrm{Sl}(2;\mathbb{R})/\mathrm{U}(1)$
with the corresponding K\"ahler potential $e^{\mathcal{K}}\ =\ \mathrm{Im}^{3}\left( Z\right)$; observe that this implies the constraint
$\mathrm{Im}\left( Z\right) > 0$.
Ignoring the possibility of gauging isometries of the resulting scalar-manifold, so that $\mathtt{P}=0$, we can calculate the potential
in eq.~(\ref{eq:Potential}) only to find
\begin{equation}
  \label{eq:25}
  \mathtt{V} \; =\; \textstyle{2g^{2}\over 3}\left[\
                       \mathtt{C}_{1}^{2}\ \mathrm{Im}^{-1}(Z) \; +\;
                       6\mathtt{C}_{0}\mathtt{C}_{1}\ \mathrm{Re}(Z)\mathrm{Im}^{-3}(Z)
                    \ \right] \; .
\end{equation}
There are two interesting sub-classes to be considered, the first one being $\mathtt{C}_{\Lambda}=(0,\mathtt{C}_{1})$
for which the potential is of the correct form to correspond to the dimensionally reduced version of the 
theory considered in \cite{Grover:2008jr}.
\par
The second case is $\mathtt{C}_{\Lambda}=(\mathtt{C}_{0},0)$, which seeing that the potential is linear in $\mathtt{C}_{0}$ means
that the potential vanishes. By construction this not only means that we can construct non-BPS solutions to the 
4-dimensional supergravity theory, but also that it can be oxidised to minimal 5-dimensional sugra. 
A simple time-like static solution for this latter case can be found by putting $\mathcal{I}^{0}=0$, so that we can take the 
base-space to be $\mathbb{R}^{3}$, and $\mathcal{I}_{1}=0$ as to ensure staticity, {\em i.e.\/} $\omega = 0$; the regularity of the solution to the stabilisation
equations, or equivalently the consistency of the metrical factor $|X|^{2}$, imposes the constraint $\mathcal{I}_{0}\left(\mathcal{I}^{1}\right)^{3}<0$.
With this information the solution is determined by
\begin{equation}
  \label{eq:32}
  \frac{1}{2|X|^{2}} \ =\ \sqrt{\ 2\ \left| \mathcal{I}_{0}\left(\mathcal{I}^{1}\right)^{3}\right|\ } 
  \hspace{.4cm} ,\hspace{.4cm}
  Z \, =\, 2i\sqrt{\ \left|\frac{\mathcal{I}_{0}}{\mathcal{I}^{1}}\right|\ } \; ,
\end{equation}
so that the solution is asymptotically Kasner. 
As the effective radius of the compactified fifth direction is proportional to $\mathrm{Im}(Z)$ which grows linear in $\tau$, this solutions is asymptotically decompactifying;
the resulting 5-dimensional metric is readily found to be (shifting $\mathcal{I}^{1}\rightarrow \sqrt{2}\ H$)
\begin{equation}
  \label{eq:33}
   ds_{(5)}^{2} \ =\ 2H^{-1}\ dy\left(\ d\tau \ -\ 2\sqrt{2}\ \left|\mathcal{I}_{0}\right|\ dy\right) \ -\ H^{2}\ d\vec{x}^{2}\; .
\end{equation}
which can be transformed to a Walker metric for a space of holonomy $\mathrm{Sim}(3)$ \cite{art:walker1950}.
Observe that the relation between $d+1$ dimensional spaces of holonomy in $\mathrm{Sim}(d-1)$ and time-dependent black holes, of which the foregoing
is one example, was first introduced and used in ref.~\cite{Gibbons:2007zu}.
\par
The generic solution in section (\ref{sec:Holomorphic}) can readily be adapted to the model at hand and reads
\begin{equation}
  \label{eq:34}
  ds^{2} \;  =\;  2du\left( dv \ +\ \lambda^{2}\ v^{2}\ \mathcal{Z}^{-3}\ du\right) \; -\; \textstyle{2\over \lambda^{2}}\ \mathcal{Z}^{3}\ \frac{dzd\bar{z}}{(1-|z|^{2})^{2}} \; , 
\end{equation}
where we introduced the abbreviations $\sqrt{2}\lambda = g\mathtt{C}_{0}$ and $\mathcal{Z}=\mathrm{Im}(Z)$.  The vector fields are given by the expression 
(\ref{eq:7d}), with $\tilde{A}^{0}=0$ and $\tilde{A}^{1}$ needs to satisfy
\begin{equation}
  \label{eq:35}
  d\tilde{A}^{1} \; =\; \textstyle{\sqrt{2}i}{\lambda}\ \mathcal{Z}\ \frac{dz\wedge d\bar{z}}{(1+|z|^{2})^{2}} \; , 
\end{equation}
which presupposes knowing the explicit dependence of $Z$ on $z$. 
\par
Lifting this solution up to 5 dimensions we obtain, after the coordinate transformations $v\rightarrow e^{\sqrt{2}\lambda y}w$ where $y$ is the 5$^{th}$ direction, 
the following solution 
\begin{eqnarray}
  \label{eq:36}
  ds^{2}_{(5)} & =& 2\mathcal{Z}^{-1}\ e^{\sqrt{2}\lambda y}\ dudw \; -\; \mathcal{Z}^{2}\left[
                                    dy^{2} \ +\ \textstyle{2\over \lambda^{2}}\ \frac{dzd\bar{z}}{(1-|z|^{2})^{2}} 
                              \right] \; , \\
  \hat{A} & =& \sqrt{3}\ \mathrm{Re}\left( Z\right)\ \left[
                               dy \ +\ 2\sqrt{2}\lambda\ \mathcal{Z}^{-3}\ vdu
                        \right] \, -\, \sqrt{3}\ \tilde{A}^{1} \; ,
\end{eqnarray}
where $\tilde{A}^{1}$ is determined by the condition (\ref{eq:36}):
this solution is  a deformation of the maximally supersymmetric $aDS_{3}\times S^{2}$ solution, and deformations of the other maximally supersymmetric
5-dimensional solutions can be obtained by using the $Sp(2;\mathbb{R})$-duality transformations before oxidation, similar to how the 4- and 5-dimensional
vacua are related (see {\em e.g.\/} ref.~\cite{LozanoTellechea:2002pn}).
\par
Let us end this section by pointing out that there are more models for which the FI-contribution to the potential vanishes \cite{Andrianopoli:1996cm}.
One of them is the $\mathcal{ST}[2,m]$-model, which in the ungauged supergravity model, allows for the embeddings of monopoles and 
the construction of non-Abelian black holes \cite{Hubscher:2008yz}, and we will briefly talk about the solutions. 
\par
A convenient parameterisation of the model is given by the symplectic section
\begin{equation}
  \label{eq:38}
  \mathcal{V} \, =\, \left(\begin{array}{c} \mathrm{L}^{\Lambda} \\ \eta_{\Lambda\Sigma}\mathrm{S}\ \mathrm{L}^{\Sigma}\end{array}\right)
  \;\;\mbox{with}\;\;
  \left\{\begin{array}{lcl}
       \eta & =& \mathrm{diag}([+]^{2},[-]^{m}) \\
        & & \\
       0     & =& \eta_{\Lambda\Sigma} \mathrm{L}^{\Lambda}\mathrm{L}^{\Sigma}
  \end{array}\right.   \; .
\end{equation}
The FI-part of the potential is easily calculated and gives \cite{Cremmer:1984hj,Andrianopoli:1996cm}
\begin{equation}
  \label{eq:39}
  \mathtt{V}_{FI} \; =\; -\textstyle{g^{2}\over 4}\ \mathrm{Im}^{-1}\left(\mathrm{S}\right)\ \mathtt{C}_{\Lambda}\eta^{\Lambda\Sigma}\mathtt{C}_{\Sigma} \; ,
\end{equation}
so that $\mathtt{V}_{FI}=0$ whenever $\mathtt{C}$ is a null-vector w.r.t.~$\eta$. Taking $\mathcal{ST}[2,4]$ as the model to work with and $\mathtt{C}$
to be a null-vector, we can gauge an $SU(2)$-gauge group, and by further taking $\mathtt{C}_{\Lambda}\mathcal{I}^{\Lambda}=0$, implying that the 
base-space is $\mathbb{R}^{3}$, we can generalise the solutions found in ref.~\cite{Meessen:2008kb} to cosmological solutions.
For that take the indices $\Lambda$ to run over $(0,+,-, i)$  (with $0$ a time-like direction, $\pm$ the null directions and $i=1,2,3$) and
let $\mathtt{C}_{+}$ be the only non-vanishing element of the $\mathtt{C}$s.
By taking then $\mathcal{I}^{\pm}=\mathcal{I}_{0}=\mathcal{I}_{i}=0$ we find a static solution, {\em i.e.\/} $\omega =0$, which allows for the 
embedding of an 't Hooft-Polyakov monopole, say, in the $\mathcal{I}^{i}$s. If we then further take $\tilde{\mathcal{I}}_{+}=0$ and normalise the 
metric on constant-$\tau$ slices to be asymptotically $\mathbb{R}^{3}$, which is equivalent to taking $\tilde{\mathcal{I}}_{-}$ and $\mathcal{I}^{0}$
to be suitable constants, we see that the metric is determined through eq.~(\ref{eq:37}) and 
\begin{equation}
  \label{eq:40}
  \frac{1}{2|X|^{2}}\; =\; \sqrt{\tau}\ \sqrt{ 1\, +\ \textstyle{\mu^{2}\over g^{2}}\left[ 1-\overline{H}^{2}\right] } \; , 
\end{equation}
where $\overline{H}$ is a completely regular function of $r\in\mathbb{R}$ coming from the 't Hooft-Polyakov monopole: it reads
\begin{equation}
  \label{eq:41}
  \overline{H} \; =\; \coth \left(\mu r\right) \ -\ \frac{1}{\mu r} \; ,
\end{equation}
and is a monotonic functions with $\overline{H}(r=0)=0$ and asymptoting to $\overline{H}(r\rightarrow\infty )=1$. This means that the 
constant-$\tau$ slices are complete: the full metric, however, suffers from an initial singularity at $\tau =0$ and also from 
Kasner expansion.
\par
More general solutions can of course be constructed by considering the hairy or coloured solutions in refs.~\cite{Meessen:2008kb,Hubscher:2008yz},
in case one is interested in non-Abelian solutions, or the general Abelian solutions of ref.~\cite{Behrndt:1997ny}; to these solutions the general the
comments made in section (\ref{sec:CosmMon}) apply.
%%%%%%%%%%%%%%%%%%%%%%%%%%%%%%%%%%%%%%%%%%%%%%%%%%%%%%%%%%%%%%%%%%%%%%%%%%%%%%%%%%%%%%%%%%%%%%%%%%%%%%%%%%%%%%%%%
\section{Conclusions...}
\label{sec:Concl}
%%%%%%%%%%%%%
In this article we studied the fake-supersymmetric solution that can be obtained from $N=2$ $d=4$ gauged supergravity coupled to (non-Abelian) vector
multiplets, by Wick-rotating the FI-term needed in order to obtain gauged supergravity. 
As is usual in the classification of (fake-)supersymmetric solutions, the solutions are divided into two classes, denoted the time-like- and
the null-case, which are distinguished by the norm of the vector built out of the preserved Killing spinor. 
\par
In the time-like case we find that the metric is of the standard conformastationary form, appearing naturally in the supersymmetric time-like solutions,
with the difference that the metric is to have a specific time dependence; this time dependence is such that there is a natural substitution principle,
as first pointed out by Behrndt and Cveti\v{c} \cite{Behrndt:2003cx}, 
of creating solutions from the known supersymmetric solutions to $N=2$ $d=4$ supergravity
coupled to (non-Abelian) vector multiplets. 
Apart from this time-dependence, we find that the base-space must be a subclass of 3-dimensional Einstein-Weyl spaces
known as hyperCR- or Gauduchon-Tod spaces \cite{Gauduchon:1998},
and that half of the seed functions, namely the $\mathcal{I}^{\Lambda}$, 
must obey the Bogomol'nyi equation generalised to GT-spaces.
\par
In the null-case we find that the solutions must have a holonomy contained in $\mathrm{Sim}(2)$, which arguably can be considered to be
a minor detail: it was, however, shown in ref.~\cite{art:coley} that the purely gravitational solutions of this kind have rather special properties
with respect to quantum corrections, and it is not unconceivable that this holds for the more general class of solutions with $\mathrm{Sim}(2)$-holonomy
in supergravity theories, such as the one presented in section (\ref{sec:PotIsNul}).
\par
We did not develop a full-fledged characterisation of the solutions in the null-case, but instead focussed on the new characteristics induced 
by the interplay between $\mathrm{Sim}(2)$-holonomy and Special Geometry. The end result is what can be considered to be a back-reacted 
solution describing the intersection of a Nariai/Robinson-Bertotti space with a generic (stringy) cosmic string \cite{Meessen:2006tu}.
\par
The fact that the holonomy is contained in $\mathrm{Sim}(2)$ is caused by the fact that we are gauging an $\mathbb{R}$-symmetry, where-from 
one deduces that the null-vector one constructs as a bilinear of the preserved Killing spinor is gauge-covariantly constant null-vector; said differently
it is a recurrent null-vector, whence the 4-dimensional space has holonomy $\mathrm{Sim}(2)$ \cite{Gibbons:2007zu}. 
As the Wick-rotation needed to create  fake supergravities from ordinary gauged supergravities will always introduce an $\mathbb{R}$-gauging,
one might be inclined to think that fake supersymmetric solutions in the null case always have infinitesimal holonomy in $\mathfrak{sim}(d-2)$.
This is, however, only partially true.
Consider for instance the theory studied by Grover {\em et al\/} \cite{Grover:2008jr}: in that case one can see that the 
recurrency condition (\ref{eq:5}) still holds {\em but} with the Levi-Civit\`a connection replaced with a metric compatible, torsionful connection, 
where the torsion is completely anti-symmetric and proportional to the Hodge dual of the graviphoton field strength. 
As the connection is metric, the link between the recurrency relation and $\mathfrak{sim}$-holonomy going through {\em mutatis mutandis},
we see that in fake $N=1$ $d=5$ gauged supergravity theories, there is a $\mathrm{Sim}(3)$ holonomy even though in general it is not
associated to the Levi-Civit\`a connection.
\par
As was shown by Gibbons {\&} Pope in ref.~\cite{Gibbons:2007zu}, and illustrated in section (\ref{sec:PotIsNul}), time-dependent solution of the
kind found in the time-like case can be obtained by dimensional reduction of spaces with $\mathrm{Sim}$-holonomy; the solutions in the time-like
case can also be obtained from the solutions in the 5-dimensional time-like case. This strongly suggest that the ordinary hierarchy of supersymmetric solutions,
and the geometric structures appearing in them, 
to theories in $d=6$, $5$ and $4$ with eight supercharges has a fake analogue.\\[.3cm]
{\bf Note added:} Shortly after this paper appeared, Gutowski {\&} Sabra \cite{Gutowski:2009vb} published
the classification of the fake supersymmetric solutions to the minimal theory. Let us for completeness
point out that the general solution to the null-case is the Nariai solution in eq.~(\ref{eq:NullNariai})
with the substitution $g_{uu}=-2g^{2}v^{2}\rightarrow -2g^{2}v^{2} +2\Upsilon_{0}(z,\bar{z})$, with
$\partial_{z}\partial_{\bar{z}}\Upsilon_{0}=0$
%%%%%%%%%%%%%%%%%%%%%%%%%%%%%%%%%%%%%%%%%%%%%%%%%%%%%%%%%%%%%%%%%%%%%%%%%%%%%%%%%%%%%%%%%%%%%%%%%%%%%%%%%%%%%%%%
\section*{Acknowledgements}
%%%%%%%%%
This work has been supported in part by
a C.S.I.C.~scholarship JAEPre-07-00176 (AP), 
the Comunidad de Madrid grant HEPHACOS P-ESP-00346, 
by the EU Research Training Network 
\textit{Constituents, Fundamental Forces and Symmetries of the Universe} MRTN-CT-2004-005104, 
the Spanish Consolider-Ingenio 2010 program CPAN CSD2007-00042 and by the {\em Fondo Social Europeo} through an
I3P-doctores scholarship (PM). 
PM wishes to thank J.~Hartong, C.~Herdeiro, D.~Klemm, T.~Ort\'{\i}n and S.~Vaul\`a for fruitful discussions.
%%%%%%%%%%%%%%%%%%%%%%%%%%%%%%%%%%%%%%%%%%%%%%%%%%%%%%%%%%%%%%%%%%%%%%%%%%%%%%%%%
%                                                                               %
%       BEGINNING OF THE APPENDICES                                             %
%                                                                               %
%%%%%%%%%%%%%%%%%%%%%%%%%%%%%%%%%%%%%%%%%%%%%%%%%%%%%%%%%%%%%%%%%%%%%%%%%%%%%%%%%
\appendix{
%%%%%%%%%%
%%%%%    SPECIAL GEOMETRY  
%%%%%%%%%%%%%%%%%%%%%%%%%%%%%%%%%%%%%
\section{Special Geometry: the bear necessities}
\label{appsec:SpecGeom}
%%%%%%
%% You can find these things is \cite{Ceresole:1995ca,FreWonderland,Craps:1997gp}.
%% Also have a look at \cite{Ceresole:1995jg}, btw it all started with \cite{Strominger:1990pd},
%% well at least the more formal set-up... The first dealing with this stuff was \cite{deWit:1984px}.
%%% {\bf Todo esto es ligeramente pedante, pero ¿que le voy a hacer?}
The formal starting point for the definition of a Special K\"ahler manifold, lies in the definition
of a K\"ahler-Hodge manifold.\footnote{
  This appendix is meant to be concise but not exhaustive. The interested reader is kindly
  referred to ref.~\cite{Andrianopoli:1996cm} and references therein.
} 
A KH-manifold is a complex line bundle over a K\"ahler manifold $\mathcal{M}$, such that
the first, and only, Chern class of the line bundle equals the K\"ahler form. This then implies that
the exponential of the K\"ahler potential can be used as a metric on the Line bundle. Furthermore,
the connection on the line bundle is
$\mathcal{Q}= (2i)^{-1}( dz^{i}\partial_{i}\mathcal{K} - d\bar{z}^{\bar{\imath}}\partial_{\bar{i}}\mathcal{K})$.
Let us denote the line bundle by $L^{1}\rightarrow \mathcal{M}$, where the superscript is there to indicate
that the covariant derivative is $\mathfrak{D} = \nabla + i\mathcal{Q}$
\par
Consider then a flat $2(n+1)$ vector bundle $E\rightarrow\mathcal{M}$ with
structure group $Sp(n+1;\mathbb{R})$, and take a section $\mathcal{V}$ of the product bundle
$E\otimes L^{1}\rightarrow\mathcal{M}$ and its complex conjugate $\overline{\mathcal{V}}$, which
is a section of the bundle $E\otimes L^{-1}\rightarrow \mathcal{M}$.
A special K\"ahler manifold, then is a bundle $E\otimes L^{1}\rightarrow\mathcal{M}$, for which there exists
a section $\mathcal{V}$ such that
\begin{equation}
  \label{eq:SGDefFund}
  \mathcal{V} \ =\ \left(
                     \begin{array}{c}
                       \mathcal{L}^{\Lambda}\\
                       \mathcal{M}_{\Lambda}
                     \end{array}
                   \right) \;\; \rightarrow \;\;
  \left\{
     \begin{array}{lcl}
       \langle \mathcal{V}\mid\overline{\mathcal{V}}\rangle & \equiv&
                  \overline{\mathcal{L}}^{\Lambda}\mathcal{M}_{\Lambda} \ -\ \mathcal{L}^{\Lambda}\overline{\mathcal{M}}_{\Lambda}
                  \; =\; -i \\
       & & \\
       \mathfrak{D}_{\bar{\imath}}\mathcal{V} & =& 0 \; ,\\
       & & \\
       \langle\mathfrak{D}_{i}\mathcal{V}\mid\mathcal{V}\rangle & =& 0 \; .
     \end{array}
  \right.
\end{equation}
% The situation w.r.t. these basic definitions is quite obscure: According to \cite{Ceresole:1995ca} everything
% should follow from the first two, which is a tad too trivial, whilst according to \cite[(4.37)]{Craps:1997gp}
% you should impose also another constrain, after which they go on to state that this extra constraint is always
% implied by the three mentioned here. My own calculations show that the above three are necessary to
% get the job done, and so I appeal to Ockham! A last word on the matter: The system is just too trivial if one
% only allows for the first 2 constraints, methinks..
\par
By defining the objects
\begin{equation}
  \label{eq:SGDefU}
  \mathcal{U}_{i} \; \equiv\; \mathfrak{D}_{i}\mathcal{V} \; =\;
       \left(
         \begin{array}{c}
             f_{i}^{\Lambda}\\
             h_{\scriptscriptstyle{\Lambda}\ i}
         \end{array}
       \right) \;\; ,\;\; \overline{\mathcal{U}}_{\bar{\imath}} \; =\; \overline{\mathcal{U}_{i}} \; ,
\end{equation}
it follows from the basic definitions that
\begin{equation}
  \label{eq:SGProp1}
  \begin{array}{lclclcl}
     \mathfrak{D}_{\bar{\imath}}\ \mathcal{U}_{i} & =& \mathcal{G}_{i\bar{\imath}}\ \mathcal{V} &\hspace{.3cm},\hspace{.3cm}&
     \langle\mathcal{U}_{i}\mid\overline{\mathcal{U}}_{\bar{\imath}}\rangle & =& i\mathcal{G}_{i\bar{\imath}} \; , \\
     & & & & & & \\
     \langle\mathcal{U}_{i}\mid\overline{\mathcal{V}}\rangle & =& 0 & ,&
     \langle\mathcal{U}_{i}\mid\mathcal{V}\rangle & =& 0 \; .
  \end{array}
\end{equation}
Let us have a look at $\langle\mathfrak{D}_{i}\mathcal{U}_{j}\mid\mathcal{V}\rangle = -\langle\ \mathcal{U}_{j}\mid \mathcal{U}_{i}\rangle$,
where we have made use of the third constraint. As one can see the r.h.s. is antisymmetric in $i$ and $j$, whereas the l.h.s.
is symmetric 
This then means that $\langle\mathfrak{D}_{i}\mathcal{U}_{j}\mid\mathcal{V}\rangle = \langle \mathcal{U}_{j}\mid \mathcal{U}_{i}\rangle = 0$.
The importance of this last equation is that if we group together $\mathcal{E}_{\Lambda} = (\mathcal{V},\mathcal{U}_{i})$, then
we can see that $\langle \mathcal{E}_{\Sigma}\mid\overline{\mathcal{E}}_{\Lambda}\rangle$ is a non-degenerate matrix, which allows
us to construct an identity operator for the symplectic indices, such that for a given section of $\mathcal{A}\ni\Gamma\left( E,\mathcal{M}\right)$
we have
\begin{equation}
  \label{eq:SGSymplProj}
  \mathcal{A} \ =\ i\langle\mathcal{A}\mid\overline{\mathcal{V}}\rangle\ \mathcal{V}
              \ -\ i\langle\mathcal{A}\mid\mathcal{V}\rangle\ \overline{\mathcal{V}}
              \ +\ i\langle\mathcal{A}\mid\mathcal{U}_{i}\rangle\mathcal{G}^{i\bar{\imath}}\ \overline{\mathcal{U}}_{\bar{\imath}}
              \ -\ i\langle\mathcal{A}\mid\overline{\mathcal{U}}_{\bar{\imath}}\rangle\mathcal{G}^{i\bar{\imath}}\mathcal{U}_{i} \; .
\end{equation}
We saw that $\mathfrak{D}_{i}\mathcal{U}_{j}$ is symmetric in $i$ and $j$, but what more can be said about it?
As one can easily see, the innerproduct with $\overline{\mathcal{V}}$ and $\overline{\mathcal{U}}_{\bar{\imath}}$ vanishes
due to the basic properties. Let us then define the weight $(2,-2)$ object
\begin{equation}
  \label{eq:SGDefC}
  \mathcal{C}_{ijk} \; \equiv\; \langle \mathfrak{D}_{i}\ \mathcal{U}_{j}\mid \mathcal{U}_{k}\rangle \;\; \rightarrow\;\;
  \mathfrak{D}_{i}\ \mathcal{U}_{j} \; =\; i\mathcal{C}_{ijk}\mathcal{G}^{k\bar{l}}\overline{\mathcal{U}}_{\bar{l}} \; ,
\end{equation}
the last equation being a consequence of eq. (\ref{eq:SGSymplProj}). Since the $\mathcal{U}$'s are orthogonal, however,
one can see that $\mathcal{C}$ is completely symmetric in its 3 indices, and 2 small calculations show that
\begin{equation}
  \label{eq:SGCProp}
  \mathfrak{D}_{\bar{\imath}}\ \mathcal{C}_{jkl} \; =\; 0 \;\; ,\;\;
  \mathfrak{D}_{[i}\ \mathcal{C}_{j]kl} \; =\; 0\; .
\end{equation}
Let us then introduce the concept of a monodromy matrix $\mathcal{N}$, which can be defined through the relations
\begin{equation}
  \label{eq:SGDefN}
  \mathcal{M}_{\Lambda} \; =\; \mathcal{N}_{\Lambda\Sigma}\ \mathcal{L}^{\Sigma} \;\; ,\;\;
  h_{\Lambda i} \; =\; \overline{\mathcal{N}}_{\Lambda\Sigma}\ f_{i}^{\Sigma} \; ,
\end{equation}
The relations of $\langle\mathcal{U}_{i}\mid\overline{\mathcal{V}}\rangle =0$ then implies that
$\mathcal{N}$ is a symmetric matrix, 
which then automatically trivialises $\langle\mathcal{U}_{i}\mid\mathcal{U}_{j}\rangle =0$.
\par
Observe that as $\mathrm{Im}\left(\mathcal{N}_{\Lambda\Sigma}\right)\equiv\mathrm{Im}\left(\mathcal{N}\right)_{\Lambda\Sigma}$
appears in the kinetic term of the ($\bar{n}=n+1$) vector fields it has to be negative definite, whence also invertible, 
in order for the 
kinetic term to be well-defined: one can see that this is implied by the properties of special 
geometry \cite{Cremmer:1984hj}. 
As it is invertible, we can use it as a `metric' for raising and lowering the $\Lambda$-indices, {\em e.g.\/}
$\mathcal{L}^{\Lambda}\equiv\mathrm{Im}\left(\mathcal{N}\right)^{-1|\Lambda\Sigma}\mathcal{L}_{\Sigma}$.
Likewise we can, and shall, use $\mathcal{G}_{i\bar{\jmath}}$ to raise and lower K\"ahler indices.
\par
From the other basic properties in (\ref{eq:SGProp1}) we find
\begin{equation}
  \label{eq:SGProp1N}
  \mathcal{L}_{\Lambda}\overline{\mathcal{L}}^{\Lambda}\ =\ -\textstyle{1\over 2} \;\; ,\;\;
  \mathcal{L}_{\Lambda}\ f^{\Lambda}_{i} \ =\ 0 \;\; ,\;\; 
  f_{\Lambda i}\ \bar{f}^{\Lambda}_{\bar{\jmath}} \ =\ -\textstyle{1\over 2}\mathcal{G}_{i\bar{\jmath}} \; .
\end{equation}
An important identity that one can derive, is given by
\begin{equation}
  \label{eq:SGImpId}
  U^{\Lambda\Sigma} \; \equiv\;  f_{i}^{\Lambda}\mathcal{G}^{i\bar{\imath}}\bar{f}^{\Sigma}_{\bar{\imath}}
                    \; =\; -\textstyle{1\over 2}\mathrm{Im}(\mathcal{N})^{-1|\Lambda\Sigma}
                    \  -\  \overline{\mathcal{L}}^{\Lambda}\mathcal{L}^{\Sigma} \; ,
\end{equation}
so that $\overline{U^{\Lambda\Sigma}}=U^{\Sigma\Lambda}$.
\par
Let us construct the $(n+1)\times (n+1)$-matrices $M=(\mathcal{M}_{\Lambda},\bar{h}_{\Lambda\ \bar{\imath}})$
and $L=(\mathcal{L}^{\Lambda}, \bar{f}^{\Lambda}_{\bar{\imath}})$. With it we can write the defining relations
for the monodromy matrix as $M_{\Lambda\Sigma} = \mathcal{N}_{\Lambda\Omega}L^{\Omega}{}_{\Sigma}$, a system
which we can easily solve by putting $\mathcal{N}=ML^{-1}$, where $L^{-1}$ is the inverse of $L$.
{}Formally one finds
\begin{equation}
  \label{eq:SGLinverse}
  L^{-1} \; =\; -2 \left(
                 \begin{array}{c}
                   \mathcal{L}_{\Lambda}\\
                   f^{\bar{\imath}}_{\Lambda}
                 \end{array}
                \right) \; ,
\end{equation}
which is a recursive argument, but is useful to derive
\begin{equation}
  \label{eq:SGdifiN}
  \partial_{\bar{\imath}}\overline{\mathcal{N}}_{\Lambda\Sigma} \; =\;
     -4i\ \left(
           \bar{f}_{\Lambda\bar{\imath}}\mathcal{L}_{\Sigma} \ +\ \mathcal{L}_{\Lambda}\bar{f}_{\Sigma\bar{\imath}}
       \right) \; ,
\end{equation}
and
\begin{equation}
  \label{eq:SGdifbiN}
  \partial_{\bar{\imath}}\mathcal{N}_{\Lambda\Sigma} \; =\;
      4\ \overline{\mathcal{C}}_{\bar{\imath}\bar{\jmath}\bar{k}}\ f^{\bar{\jmath}}_{\Lambda}\ f^{\bar{k}}_{\Sigma} \; .
\end{equation}
These last equations are used extensively in the null-case.
\par
%%%%%%%%%
In explicit constructions of the models it is worthwhile to introduce the explicitly holomorphic section 
$\Omega = e^{-\mathcal{K}/2}\mathcal{V}$, which allows us to rewrite
the system (\ref{eq:SGDefFund}) as
\begin{equation}
  \label{eq:SGDefFund2}
  \Omega \ =\ \left(
                     \begin{array}{c}
                       \mathcal{X}^{\Lambda}\\
                       \mathcal{F}_{\Sigma}
                     \end{array}
                   \right) \;\; \rightarrow \;\;
  \left\{
     \begin{array}{lcl}
       \langle \Omega \mid\overline{\Omega}\rangle & \equiv&
                  \overline{\mathcal{X}}^{\Lambda}\mathcal{F}_{\Lambda} \ -\ \mathcal{X}^{\Lambda}\overline{\mathcal{F}}_{\Lambda}
                  \; =\; -i\ e^{-\mathcal{K}} \\
       & & \\
       \partial_{\bar{\imath}}\Omega & =& 0 \; ,\\
       & & \\
       \langle\partial_{i}\Omega\mid\Omega\rangle & =& 0 \; .
     \end{array}
  \right.
\end{equation}
If we now assume that $\mathcal{F}_{\Lambda}$ depends on $Z^{i}$ through the $\mathcal{X}$'s, then from the last equation we
can derive that
\begin{equation}
  \partial_{i}\mathcal{X}^{\Lambda}\left[
       2\mathcal{F}_{\Lambda} \ -\ \partial_{\Lambda}\left( \mathcal{X}^{\Sigma}\mathcal{F}_{\Sigma}\right)
  \right] \; =\; 0 \; .
\end{equation}
If $\partial_{i}\mathcal{X}^{\Lambda}$ is invertible as a $n\times (n+1)$ matrix, then we must conclude that
\begin{equation}
  \label{eq:Prepot}
  \mathcal{F}_{\Lambda} \; =\; \partial_{\Lambda}\mathcal{F}(\mathcal{X}) \; ,
\end{equation}
where $\mathcal{F}$ is a homogeneous function of degree 2, baptised by the literature as the {\em prepotential}.
\par
Making use of the prepotential and the definitions (\ref{eq:SGDefN}), we can then calculate
\begin{equation}
  \label{eq:PrepotN}
  \mathcal{N}_{\Lambda\Sigma} \; =\; \overline{\mathcal{F}}_{\Lambda\Sigma}
                  \ +\ 2i\textstyle{
                    \mathrm{Im}(\mathcal{F})_{\Lambda\Lambda^{\prime}}\mathcal{X}^{\Lambda^{\prime}}
                     \mathrm{Im}(\mathcal{F})_{\Sigma\Sigma^{\prime}}\mathcal{X}^{\Sigma^{\prime}}
                               \over
                      \mathcal{X}^{\Omega}\mathrm{Im}(\mathcal{F})_{\Omega\Omega^{\prime}}\mathcal{X}^{\Omega^{\prime}}
                                         }\; ,
\end{equation}
which, though not beautiful, is at least manifestly symmetric.
From the above expression we can obtain the sometimes useful result
\begin{equation}
  \label{eq:PrePotImNinv}
  \mathrm{Im}\left(\mathcal{N}\right)^{-1|\Lambda\Sigma} \; =\; 
        -\mathrm{F}^{-1|\Lambda\Sigma}
        \ -\ 2\mathcal{L}^{\Lambda}\overline{\mathcal{L}}^{\Sigma}
        \ -\ 2\overline{\mathcal{L}}^{\Lambda}\mathcal{L}^{\Sigma} \; ,
\end{equation}
where $\mathrm{F}^{-1}$ is the inverse of 
$\mathrm{F}_{\Lambda\Sigma}\equiv\mathrm{Im}\left(\mathcal{F}_{\Lambda\Sigma}\right)$.
Having the explicit form of $\mathcal{N}$ we can derive an explicit representation for $\mathcal{C}$,
namely
\begin{equation}
  \label{eq:PrepC}
  \mathcal{C}_{ijk} \; =\; e^{\mathcal{K}}\
                           \partial_{i}\mathcal{X}^{\Lambda}\ \partial_{j}\mathcal{X}^{\Sigma}\
                           \partial_{k}\mathcal{X}^{\Omega}\
                           \mathcal{F}_{\Lambda\Sigma\Omega} \; ,
\end{equation}
so that the prepotential determines all structures in special geometry.
%%%%%%%%%%%%%%%%%%%%%%%%%%%%%%%%%%%%%%%%%%%%%%%%%%%%%%%%%%%%%%%%%%%%%%%%%%%%%%%%%%%%%%%%%%%%%%%%%%%%%%%%
\subsection{Killing vectors in special geometry}
\label{sec:SGisom}
%%%%%%%%
We are interested in holomorphic Killing vectors associated to the K\"ahler manifold with metric $\mathcal{G}$.
More to the point, we consider the real vector
\begin{equation}
  \label{eq:SGK1}
  \mathtt{K} \; =\; \mathtt{K}^{i}(Z)\ \partial_{i} \; +\; \bar{\mathtt{K}}^{\bar{\imath}}(\overline{Z})\ \partial_{\bar{\imath}}
      \;\; \longrightarrow\;\;  \pounds_{\mathtt{K}}\mathcal{G} \; =\; 0 \; .
\end{equation}
For reasons that have to do with the number of available vectors in the theory, $\bar{n}\equiv n+1$, we can
only use $\bar{n}$ of the possible Killing vectors, and therefore we shall always label the Killing vectors
by an index like $\Lambda$, even though we are not going to use all $\bar{n}$ of them;
in fact, as we have to use 1 gauge field to gauge the $\mathbb{R}$-symmetry, we can use at most $n$
vectors to gauge isometries. 
\par
In general these
Killing vectors define a non-Abelian algebra, which we take to be
\begin{equation}
  \label{eq:SGK2}
  \left[\ \mathtt{K}_{\Lambda}\ ,\ \mathtt{K}_{\Sigma}\ \right] \; =\; 
   -\mathtt{f}_{\Lambda\Sigma}{}^{\Gamma}\ \mathtt{K}_{\Gamma} \; .
\end{equation}
\par
The isometries need not leave invariant the K\"ahler potential, in stead they must leave it invariant up to a 
K\"ahler transformation, {\em i.e.\/}
\begin{equation}
  \label{eq:SGK3}
  \pounds_{\Lambda}\mathcal{K} \; \equiv\; \mathtt{K}_{\Lambda}\ \mathcal{K} \; =\;
          \lambda_{\Lambda}(Z) \; +\; \overline{\lambda_{\Lambda}(Z)} \; , 
\end{equation}
where we used the conventions that by $\pounds_{\Lambda}$ we actually mean $\pounds_{\mathtt{K}_{\Lambda}}$.
It is clear that the K\"ahler transformation parameters $\lambda$ have to form a representation under
the group that we are gauging and in fact one sees that
\begin{equation}
  \label{eq:SGK4}
  \pounds_{\Lambda}\lambda_{\Sigma} \, -\, \pounds_{\Sigma}\lambda_{\Lambda}
  \; =\; -\mathtt{f}_{\Lambda\Sigma}{}^{\Omega}\ \lambda_{\Omega} \; .
\end{equation}
\par
If we then also assume that the Killing vectors are compatible with the complex structure $\mathcal{J}$ defined
on the K\"ahler manifold, and therefore also with the K\"ahler form $\mathsf{K}(X,Y)\sim \mathcal{G}(\mathcal{J}X,Y)$, 
we can derive
\begin{equation}\label{eq:SGK4a}
  \pounds_{\Lambda}\ \mathsf{K} \; =\; d\left( \imath_{\Lambda}K\right) \;\; \longrightarrow\;\;
  2\pi\ \imath_{\Lambda}K \; =\; d\mathtt{P}_{\Lambda} \; ,  
\end{equation}
where the object $\mathtt{P}_{\Lambda}$ is called the {\em momentum map associated to $\mathtt{K}_{\Lambda}$}.
A closed form for the momentum map can be easily seen to be 
\begin{equation}
  \label{eq:SGK5}
  i\mathtt{P}_{\Lambda} \; =\; \textstyle{1\over 2}\left(
             \mathtt{K}_{\Lambda}^{i}\ \partial_{i}\mathcal{K} \, -\,
             \mathtt{K}_{\Lambda}^{\bar{\imath}}\ \partial_{\bar{\imath}}\mathcal{K}
                              \ -\ \lambda_{\Lambda} \ +\ \overline{\lambda}_{\Lambda}
           \right)
    \; =\; \mathtt{K}_{\Lambda}^{i}\ \partial_{i}\mathcal{K}
                              \ -\ \lambda_{\Lambda} \; , 
\end{equation}
where we made use of eq. (\ref{eq:SGK3}) and fixed a possible constant to be zero. 
Using this form and eq. (\ref{eq:SGK4}), it is straightforward to show that
\begin{equation}
  \label{eq:SGK6}
  \pounds_{\Lambda}\mathtt{P}_{\Sigma} \; =\; -\mathtt{f}_{\Lambda\Sigma}{}^{\Omega}\ \mathtt{P}_{\Omega}\; ,
\end{equation}
\par
The action of the Killing vector on the symplectic section is most easily described on the $(1,0)$-weight
section $\Omega$. In fact, by consistency it must transforms as
\begin{equation}
  \label{eq:SGK7}
  \pounds_{\Lambda}\ \Omega \; =\; S_{\Lambda}\ \Omega \, -\, \lambda_{\Lambda}\ \Omega \, ,
\end{equation}
where $S\in\mathfrak{sp}(\bar{n};\mathbb{R})$ and forms a representation of the algebra we are gauging,
{\em i.e.\/} $[S_{\Lambda},S_{\Sigma}]=\mathtt{f}_{\Lambda\Sigma}{}^{\Gamma}\ S_{\Gamma}$.
The natural space-time, not K\"ahler, connection that acts on this symplectic section is
\begin{equation}
  \label{eq:SGK8}
  \mathtt{D}\Omega \; =\; \left( \nabla \ +\
                  \partial Z^{i}\ \partial_{i}\mathcal{K} \ +\ 
                  i\mathtt{g}\ A^{\Lambda}\ \mathtt{P}_{\Lambda} \ +\
                   \mathtt{g}\ A^{\Lambda}\ S_{\Lambda}
                \right)\ \Omega \; ,
\end{equation}
which is constructed in such a way that 
$\delta_{\alpha}\mathtt{D}\Omega = \alpha^{\Lambda}\left( S_{\Lambda} -\lambda_{\Lambda}\right)\mathtt{D}\Omega$.
From the above equation it is a small calculation to derive the covariant derivative on objects such
as $\mathcal{V}$ or $\overline{\mathcal{V}}$. In fact, one can see that if we are dealing with a 
symplectic $(p,q)$-weight object, then we have
\begin{eqnarray}
  \label{eq:SGK9}
  \delta_{\alpha}\Phi^{(p,q)} & =& \alpha^{\Lambda}\left(\ 
        S_{\Lambda} \ -\ p\ \lambda_{\Lambda}\ -\ q\ \bar{\lambda}_{\Lambda}
        \right)\ \Phi^{(p,q)} \hspace{4cm} {\&} \nonumber \\
  & & \nonumber\\
  \mathtt{D}\Phi^{(p,q)} & =& \left[\nabla\ +\
            p\ \partial Z^{i}\ \partial_{i}\mathcal{K}\ +\
            q\ \partial\overline{Z}^{\bar{\imath}}\ \partial_{\bar{\imath}}\mathcal{K} \ +\
            i(p-q)\mathtt{g}\ A^{\Lambda}\ \mathtt{P}_{\Lambda} \ +\
            \mathtt{g}\ A^{\Lambda}\ S_{\Lambda}
          \right]\ \Phi^{(p,q)} \; \longrightarrow \nonumber \\
  & & \nonumber \\
  \delta_{\alpha}\mathtt{D}\Phi^{(p,q)} & =& \alpha^{\Lambda}\left(\ 
        S_{\Lambda} \ -\ p\ \lambda_{\Lambda}\ -\ q\ \bar{\lambda}_{\Lambda}
        \right)\ \mathtt{D}\Phi^{(p,q)} \; .
\end{eqnarray}
Now that we have defined the various covariant derivatives, we can go on to derive
\begin{equation}
  \label{eq:SGK10}
  \mathtt{K}^{i}\ \mathcal{U}_{i} \; =\; \left( S_{\mathtt{K}} \ +\ i\mathtt{P}_{\mathtt{K}}\right)\ \mathcal{V}
  \;\longrightarrow\;
  \mathtt{D}\mathcal{V} \; =\; \mathtt{D}Z^{i}\ \mathcal{U}_{i} \; ,
\end{equation}
which in its turn can be used to obtain
\begin{equation}
  \label{eq:SGK11}
  \mathtt{D}\mathcal{U}_{i} \; =\; \mathtt{D}Z^{j}\ \mathfrak{D}_{j}\mathcal{U}_{i} \; +\;
                    \mathtt{D}\overline{Z}^{\bar{\jmath}}\ \mathfrak{D}_{\bar{\jmath}}\mathcal{U}_{i} 
  \;\;\;\mbox{and}\;\;\;
   \mathtt{D}\mathcal{N}\; =\; \mathtt{D}Z^{i}\ \partial_{i}\mathcal{N}\; +\;
   \mathtt{D}\overline{Z}^{\bar{\imath}}\ \partial_{\bar{\imath}}\mathcal{N}\; .
\end{equation}
Equation (\ref{eq:SGK10}) allows us to write down the following identities
\begin{equation}
  \label{eq:SGK12}
  \begin{array}{lclclcl}
    0 & =& \langle \mathcal{V}\mid S_{\Lambda}\mathcal{V}\rangle &\hspace{.4cm},\hspace{.4cm}&
    \mathtt{P}_{\Lambda} & =& \langle \overline{\mathcal{V}}\mid S_{\Lambda}\mathcal{V}\rangle \; ,\\
      & & & & & & \\
    \mathtt{K}_{\Lambda\bar{\imath}} & =& i\langle \overline{\mathcal{U}}_{\bar{\imath}}\mid S_{\Lambda}\mathcal{V}\rangle & ,& 
    0 & =&   \langle \mathcal{U}_{i}\mid\ S_{\Lambda}\mathcal{V}\rangle \; . 
  \end{array}
\end{equation}
\par
As in ref.~\cite{Hubscher:2008yz}, we shall restrict ourselves to a subset of possible gaugings that we consider:
in fact we shall restrict ourselves to groups whose embedding into $\mathfrak{sp}(\bar{n};\mathbb{R})$
is given by
\begin{equation}
  \label{eq:SGK15}
  S_{\Lambda} \; =\; \left(
    \begin{array}{lcl}
      \left[ S_{\Lambda}\right]^{\Sigma}{}_{\Omega} & 0 \\
      0 & -\left[ S_{\Lambda}\right]_{\Sigma}{}^{\Omega}
    \end{array}
   \right) \; =\;  
   \left(
    \begin{array}{lcl}
      \mathtt{f}_{\Lambda\Omega}{}^{\Sigma} & 0 \\
      0 & -\mathtt{f}_{\Lambda\Sigma}{}^{\Omega}
    \end{array}
   \right)\;\; .
\end{equation}
With this restriction on the gaugeable symmetries, we can then derive the following important identity
\begin{equation}
  \label{eq:SGK16}
  0 \; =\; \mathcal{L}^{\Lambda}\ \mathtt{K}_{\Lambda}^{i} \; .
\end{equation}
Further identities that follow are
\begin{equation}
  \label{eq:SGK17}
   \mathcal{L}^{\Lambda}\ \mathtt{P}_{\Lambda} \ =\ 0 \hspace{.4cm} ,\hspace{.4cm}
   \mathcal{L}^{\Lambda}\ \lambda_{\Lambda} \ =\ 0 \hspace{.4cm} ,\hspace{.4cm}
   \bar{f}^{\Lambda\ i}\ \mathtt{P}_{\Lambda} \ =\ i\ \overline{\mathcal{L}}^{\Lambda}\ \mathtt{K}_{\Lambda}^{i} \; .
\end{equation}
%%%%%%%%%%%%%%%%%%%%%%%%%%%%%%%%%%%%%%%%%%%%%%%%%%%%%%%%%%%%%%%%%%%%%%%%%%%%%%%%%%%%%%%%%%%%%%%%%%%%%%%%
%%%%%%%%%%%%%%%%%%%%%%%%%%%%
%%%%% DEFINITION OF THE BILINEARS ETC.
%%%%%%%%%%%%%%%%%%%%%%%%%%%%
\section{Bilinears and Fierz identities}
\label{sec:Bil}
%%%%%%%%%%%%%%%
In this appendix we shall present the definitions of the bilinears;
the definitions used in this article are based on, but not equal to, those of ref.~\cite{Meessen:2006tu}.
\par
The scalar-bilinears are defined by
\begin{equation}
  \label{eq:Bil1}
  \begin{array}{lclclcl}
    X & =& \textstyle{1\over 2}\varepsilon^{IJ}\ \bar{\epsilon}_{I}\epsilon_{J} 
      &\hspace{.5cm},\hspace{.5cm}&
    \bar{\epsilon}_{I}\epsilon_{J} & =& \varepsilon_{IJ}\ X\; ,\\
     & & & & & &\\
    \overline{X} & =& \textstyle{1\over 2}\varepsilon_{IJ}\ \bar{\epsilon}^{I}\epsilon^{J} 
      &\hspace{.5cm},\hspace{.5cm}&
    \bar{\epsilon}^{I}\epsilon^{J} & =& \varepsilon^{IJ}\ \overline{X}\; .
  \end{array}
\end{equation}
The vector bilinears are defined by
\begin{equation}
  \label{eq:Bil2}
  V_{a}^{I}{}_{J} \;\equiv\; i\bar{\epsilon}^{I}\gamma_{a}\epsilon_{J}
                 \; =\; \textstyle{1\over 2}\ V_{a}\ \delta^{I}{}_{J} 
                 \ +\ \textstyle{1\over 2}\ V^{x}_{a}\ \left(\sigma^{x}\right)^{I}{}_{J}\; ,
\end{equation}
which can be inverted to
\begin{equation}
  \label{eq:Bil2a}
  V_{a} \; =\; V_{a}^{I}{}_{I} \hspace{1cm}\mbox{and}\hspace{1cm}
  V^{x}_{a} \; =\; \left(\sigma^{x}\right)_{I}{}^{J}\ V_{a}^{I}{}_{J} \; .
\end{equation}
{}Finally we have 3 imaginary-selfdual 2-forms defined by
\begin{equation}
  \label{eq:Bil3}
  \Phi_{IJ\ ab} \;\equiv\; \bar{\epsilon}_{I}\gamma_{ab}\epsilon_{J}
               \; =\; \Phi^{x}_{ab}\ \textstyle{i\over 2}\left(\sigma^{x}\right)_{IJ}
  \;\; \longrightarrow\;\;
  \Phi^{x} \; =\; i\left(\sigma^{x}\right)^{IJ}\ \Phi_{IJ} \; . 
\end{equation}
The anti-imaginary-self-dual 2-forms are defined by complex conjugation.
\par
{}From the Fierz identities we can then derive that 
\begin{equation}
  \label{eq:Bil4}
  \eta_{ab} \; =\; \frac{1}{4|X|^{2}}\left[
                      V_{a}V_{b} \; -\; V^{x}_{a}V^{x}_{b}
                  \right] \; ,
\end{equation}
and consistently with the above that 
\begin{equation}
  \label{eq:Bil5}
  \imath_{V}V^{x}\ =\ 0 \;\; ,\;\;
  g\left( V,V\right) \ =\ 4|X|^{2} \;\; ,\;\;
  g\left( V^{x},V^{y}\right) \ =\ -4|X|^{2}\ \delta^{xy} \; .
\end{equation}
A result that is harder to be found is 
\begin{equation}
  \label{eq:Bil6}
  \overline{X}\ \Phi^{x}_{ab}\; =\; -i\ 
      \left[
         V_{[a}V_{b]}^{x} \ +\ \textstyle{i\over 2}\varepsilon_{ab}{}^{cd}V_{c}V_{d}^{x}\
      \right] \; ,
\end{equation}
which translates to
\begin{equation}
  \label{eq:12}
  \overline{X}\ \Phi^{x} \; =\; \frac{1}{2i}\; \left[
                      V\wedge V^{x} \ +\ i\ \star\left( V\wedge V^{x}\right)
                  \right] \; ,
\end{equation}
in form notation.
\par
In the null-case, {\em i.e.\/} when $X=0$, the $V^{x}$ are proportional to $V$ and
the $\Phi$s become linear dependent, severely limiting the utility of the bilinears. 
In section (\ref{sec:Null}), we will, following ref.~\cite{Tod:1983pm}, introduce an auxiliar spinor which leads 
to Fierz identities similar to the ones above.
%%%%%%%%%%%%%%%%%%%%%%%%%%%%%%%%%%%%%%%%%%%%%%%%%%%%%%%%%%%%%%%%%%%%%%%%%%%%%%%%%
\section{Curvatures for the null case}
\label{sec:NullCurv}
%%%%%%%%%%%%%%%%%%%%
Let us set-up a null-Vierbein by
\begin{equation}
  ds^{2}_{null} \;=\; e^{+}\otimes e^{-} \ +\ e^{-}\otimes e^{+}
                     \ -\ e^{\bullet}\otimes e^{\bar{\bullet}}
                     \ -\ e^{\bar{\bullet}}\otimes e^{\bullet} \; ,
\end{equation}
and choose\footnote{
 The directional derivatives $\theta_{a}$ are normalised such that $e^{a}(\theta_{b})=\delta^{a}{}_{b}$.
}
\begin{equation}
  \label{eq:NCtetrad}
  \begin{array}{lclclclclcl}
    e^{+} & =& L & =& du 
          &\hspace{.3cm},\hspace{.3cm}& 
    \theta_{+} & =& N^{\flat} & =& \partial_{u} \ -\ H\partial_{v}\; ,\\
    e^{-} & =& N & =& dv +Hdu+ \varpi dz +\overline{\varpi}d\bar{z} 
    & ,&
    \theta_{-} & =& L^{\flat} & =&\partial_{v} \; ,\\
    e^{\bullet} & =& M & =& e^{U}dz
    & ,& 
    \theta_{\bullet} & =& -\overline{M}^{\flat}& =& e^{-U}\left[ \partial_{z}-\varpi\partial_{v}\right]\; ,\\
    e^{\bar{\bullet}} & =& \overline{M} & =&  e^{U}d\bar{z} 
    & ,&
    \theta_{\bar{\bullet}} & =& -M^{\flat} & =& e^{-U}\left[ \partial_{\bar{z}}-\overline{\varpi}\partial_{v}\right] \; , 
  \end{array}
\end{equation}
where conforming to the results of eq.~(\ref{eq:9}) only $H=H(u,v,z,\bar{z})$ and 
$U$ and the $\varpi$s depend on $u$, $z$ and $\bar{z}$.
\par
The non-vanishing components of the spin-connection are then seen to be 
\begin{eqnarray}
  \label{eq:NCspincon}
  \omega_{+-} & =& -\theta_{-}H\ e^{+} \; , \\
  \omega_{+\bullet} & =& \left( e^{-U}\theta_{+}\varpi\ -\ \theta_{\bullet}H\right)\ e^{+}
                   \ -\ \left[
                           \theta_{+}U 
                           \ +\ \textstyle{1\over 2} e^{-2U}\left(
                                \partial_{z}\overline{\varpi}-\partial_{\bar{z}}\varpi
                        \right)\right]\ e^{\bar{\bullet}} \; , \\
   \omega_{+\bar{\bullet}} & =& \left( e^{-U}\theta_{+}\overline{\varpi}\ -\ \theta_{\bar{\bullet}}H\right)\ e^{+}
                   \ -\ \left[
                           \theta_{+}U 
                           \ -\ \textstyle{1\over 2} e^{-2U}\left(
                                \partial_{z}\overline{\varpi} -\partial_{\bar{z}}\varpi
                        \right)\right]\ e^{\bullet} \; , \\
   \omega_{\bullet\bar{\bullet}} & =& 
               \textstyle{1\over 2}e^{-2U}
                        \left(
                           \partial_{z}\overline{\varpi} -\partial_{\bar{z}}\varpi
                        \right)\ e^{+}
               \ -\ e^{\bullet}\theta_{\bullet}U
               \ +\ e^{\bar{\bullet}}\ \theta_{\bar{\bullet}}U \; .
\end{eqnarray}
A further calculation then leads to the Ricci tensor, whose non-vanishing coefficients are
\begin{eqnarray}
  \label{eq:NCR+-}
  R_{+-} & =& -\theta_{-}^{2}H\; ,\\
  \label{eq:NCRzbz}
  R_{\bullet\bar{\bullet}} & =& 2e^{-2U}\ \partial_{z}\partial_{\bar{z}}U\; ,\\
  \label{eq:NCR+z}
  R_{+\bullet}  & =& e^{-U}\theta_{+}\partial_{z}U 
               \ -\ \theta_{\bullet}\theta_{-}H
               \ +\ \textstyle{1\over 2}\theta_{\bullet}\left( 
                        e^{-2U}\left[
                           \partial_{z}\overline{\varpi} -\partial_{\bar{z}}\varpi
                        \right]\right) \; , \\
  \label{eq:NCR+bz}
  R_{+\bar{\bullet}} & =& \overline{R_{+\bullet}} \; ,\\
  \label{eq:NCR++}
  R_{++} & =& 2e^{-U}\theta_{+}^{2}e^{U} 
           \ +\ 2\theta_{-}H\ \theta_{+}U
           \ +\ \textstyle{1\over 2}e^{-4U}
                \left( \partial_{z}\overline{\varpi} -\partial_{\bar{z}}\varpi\right)^{2}
         \nonumber\\
        &  & -e^{-U}\theta_{\bullet}\left[ e^{U}\theta_{\bar{\bullet}}H\right]
             -e^{-U}\theta_{\bar{\bullet}}\left[ e^{U}\theta_{\bullet}H\right]
             + e^{-2U}\partial_{u}\left(
                 \partial_{z}\overline{\varpi} +\partial_{\bar{z}}\varpi
              \right)\; .
\end{eqnarray}
Observe that the last term in eq.~(\ref{eq:NCR++}) can always be put to zero by the coordinate
transformation $v\longrightarrow v+\rho (u,z,\bar{z})$.
%%%%%%%%% 
%%%%%%%%%%%%%%%%%%%%%%%%%  END OF THE APPENDICES  %%%%%%%%%%%%%%%%%%%%%%%%%%%%%%%
}
%%%%%%%%%%%%%%%%%%%%%%%%%%%%%%%%%%%%%%%%%%%%%%%%%%%%%%%%%%%%%%%%%%%%%%%%%%%%%%%%%

%%%%%%%%%%%%%%%%%%%%%%%%%%%%%%%%%%%%%%%%%%%%%%%%%%%%%%%%%%%%%%%%%%%%%%%%%%%%%%%%%
\end{document}